\documentclass[journal]{IEEEtran}
\usepackage{cite}
\usepackage[caption=false,font=normalsize,labelfont=sf,textfont=sf]{subfig}

\usepackage{graphicx}
\usepackage{url}
\usepackage{times}
\usepackage{amsmath}   
\usepackage{amsfonts,amsthm,amssymb}
\usepackage[normalem]{ulem}
\usepackage{mathtools}
\usepackage{tablefootnote}
\usepackage{nicefrac}
\usepackage[none]{hyphenat}
\usepackage{caption}
\usepackage{fontenc}
\usepackage{color}
\usepackage{hyperref}
\usepackage{url}
\usepackage[table,xcdraw]{xcolor}
\usepackage{balance}
\usepackage{etoolbox}
\usepackage{multirow}
\newcolumntype{P}[1]{>{\centering\arraybackslash}p{#1}}
\newcolumntype{M}[1]{>{\centering\arraybackslash}m{#1}}
\usepackage{verbatim} 
\usepackage{float}
\usepackage{placeins}
\usepackage{comment}
\usepackage{gensymb}
\usepackage{dblfloatfix}
\usepackage{pifont}
\newcommand{\xmark}{\ding{55}}%

\newcommand{\argmax}[1]{\underset{#1}{\operatorname{arg}\,\operatorname{max}}\;}
\begin{document}
\title{Reconfigurable Radar Signal Processing Accelerator for Integrated Sensing and Communication System}

\author{Aakanksha Tewari*, Shragvi Sidharth Jha*,  Akanksha~Sneh, Sumit~J~Darak~\IEEEmembership{Senior Member~IEEE} and Shobha~Sundar~Ram,~\IEEEmembership{Senior Member~IEEE}\thanks{*Aakanksha Tewari and Shragvi Sidharth Jha are joint first-authors.}  
\thanks{All authors are with the Indraprastha Institute of Information Technology Delhi, New Delhi 110020 India. E-mail: \{aakankshat,shragvi19207,sumit, shobha,akankshas\}@iiitd.ac.in.}%
\thanks{This work is supported by the funding received from MeiTy, Chip to Startup (C2S), Government of India (Project no.: EE-9/2/2021-R\&D-E), and Qualcomm Innovation Fellowship 2022 India.}
}

\maketitle
\begin{abstract}
IEEE 802.11ad-based integrated sensing and communications (ISAC) have been identified as a potential solution for enabling next-generation intelligent transportation systems in the millimeter wave (mmW) spectrum. 
The radar functionality within the ISAC enables accurate detection and localization of mobile targets, which can significantly speed up the selection of the optimal high-directional narrow beam required for mmW communications between the base station and mobile target.  To bring ISAC to reality, a radar signal processing (RSP) accelerator, co-located with the wireless communication physical layer (PHY), on edge platforms, is desired. In this work, we discuss the three-dimensional digital hardware RSP framework for 802.11ad-based ISAC to detect the range, azimuth, and Doppler velocity of multiple targets. We present an efficient reconfigurable architecture for RSP on multi-processor system-on-chip (MPSoC) via hardware-software co-design, word-length optimization, and serial-parallel configurations.  We demonstrate the functional correctness of the proposed fixed-point architecture and significant savings in resource utilization ($\sim 40-70\%$), execution time ($1.5 \times$ improvement), and power consumption (50\%) over floating-point architecture. The acceleration on hardware offers a 120-factor improvement in execution time over the benchmark Quad-core processor. The proposed architecture enables on-the-fly reconfigurability to support different azimuth precision and Doppler velocity resolution, offering a real-time trade-off between functional accuracy and detection time. We implement end-to-end ISAC comprising RSP and PHY on MPSoC and demonstrate significant improvement in throughput over IEEE 802.11ad. 
\end{abstract}
\section{Introduction}
Integrated sensing and communications (ISAC) based dual-functional joint radar and communication waveforms, algorithms, and prototypes are being researched in response to spectral congestion issues. They can be broadly divided into research focusing on interference management between independently developed radar and communication systems 
\cite{li2017joint,martone2017spectrum,hessar2016spectrum,qian2020radar}; opportunistic exploitation of radar transmission for secondary communication applications \cite{hessar2016spectrum}; and vice versa \cite{mishra2019cognitive,kumari_ieee_2018,duggal2020Doppler}; and collaborative design of radar and communications functionalities using commonly shared waveform and hardware platform \cite{Hassanien2019dual,ram2022optimization}. The proposed work in this paper falls under this third category, where we focus on a novel digital hardware prototype implementation of a radar signal processing (RSP) accelerator for ISAC comprising of co-designed radar (for sensing) and communication for millimeter wave (mmW) vehicular communications. In literature, the term joint radar communication (JRC) system is also used for ISAC. We use both terms interchangeably. 

Over the last decade, vehicle-to-everything (V2X) communication has been researched for increased road safety and improved driver and passenger comfort for realizing semi and fully-autonomous driving. The state-of-the-art V2X communication protocols - such as the dedicated short-range communications \cite{kenney2011dedicated}, device-to-device based V2X \cite{asadi2014survey} and cellular V2X \cite{wang2018cellular} - operate at sub-6GHz carrier frequencies and hence do not support large bandwidths required for sharing time-critical high definition three-dimensional environmental sensing information. The alternative that has been suggested is high bandwidth mmW V2X communications supported by IEEE 802.11ad/ay protocols \cite{choi2016millimeter}. Interestingly, the preamble within the packet structure of these protocols has demonstrated perfect autocorrelation properties that make them attractive for radar remote sensing. Hence, there have been several recent works that have investigated the use of the IEEE 802.11ad protocol for ISAC purposes. \cite{kumari_ieee_2018,duggal2019micro,duggal2020Doppler}.
\begin{table*}[!t]
\footnotesize
 \centering
  \caption{\footnotesize Literature survey comparing the contributions of the proposed work with the prior art } 
  \renewcommand{\arraystretch}{1}
 \resizebox{\textwidth}{!}{
  \begin{tabular}{M{1.9cm}|M{0.9cm}|M{1.6cm}|M{2.1cm}|M{2cm}|M{2cm}|M{2.1cm}|M{2cm}|M{2cm}}\hline
       Reference & HSCD & RSP on Edge & Multiple Targets Detection & ISAC Waveform & ISAC Codesign & Reconfigurability & 3D Localization & {ISAC on MPSoC}\\\hline\hline
        \cite{drozdenko2017hardware} &\checkmark &\xmark&\xmark& \xmark  & \xmark &\checkmark &\xmark &\xmark
\\\hline
    \cite{agrawal2019performance}&\checkmark&\xmark&\xmark& \xmark&\xmark  &\checkmark&\xmark &{\xmark}
  \\\hline
         \cite{schweizer2021fairy}&\checkmark&\checkmark&\checkmark &\xmark&\xmark & \xmark &\checkmark     &{\xmark}
\\\hline
      \cite{zhong2016design}& \checkmark &\checkmark& \checkmark  &\xmark &\xmark &\xmark &\xmark &{\xmark}
\\\hline
    \cite{ma2021spatial}&\xmark &\xmark& \checkmark  &\checkmark &\xmark&\xmark &\xmark &{\xmark} 
\\\hline
    \cite{kumari2021jcr70}& \xmark &\checkmark& \checkmark &\checkmark &\checkmark&\xmark &\xmark&{\xmark}
\\\hline
    \cite{pegoraro2021rapid}& \xmark&\checkmark & \checkmark&\checkmark  &\checkmark &\xmark &\xmark&{\xmark}
\\\hline
    \cite{alaee2022cognitive}&\xmark&\xmark&\checkmark&\checkmark  &\xmark &\xmark &\xmark &{\xmark}
\\\hline

Proposed work  &\checkmark &\checkmark&\checkmark &\checkmark&\checkmark &\checkmark&\checkmark&{\checkmark}
\\\hline
   \end{tabular}}
   \vspace{-0.2cm}
   \label{tab:literature_review}
\end{table*}

In general, there has been significant research focus on ISAC waveform generation \cite{chiriyath2019novel,zhou2019joint,9722943} and algorithms \cite{mishra2019toward,zhang2021overview}, but limited research on experimental demonstrations of ISAC software/hardware prototypes.  In \cite{sneh2022ieee}, the architecture, design details, and software prototype of an IEEE 802.11ad-based JRC transceiver were presented.  In \cite{alaee2022cognitive}, the authors presented a hardware prototype on a universal software radio peripheral platform where they studied interference management between independently generated communication and radar waveforms. A codesigned ISAC hardware prototype based on index modulation was presented in \cite{ma2021spatial}, where the transmitting antennas were divided into radar-centric and communication-centric subarrays. ISAC transmission was experimentally demonstrated in
\cite{kumari2021jcr70} with separate radar and communications receivers. In \cite{pegoraro2021rapid}, IEEE802.11ay access points were retrofitted for indoor radar detection of humans. In these works, the RF/mmW data is downconverted, digitized, and then processed offline in programming platforms (such as MATLAB, Python, etc.) through double-precision floating-point implementations. 

For real-world deployments, the RSP must be realized on the edge platforms (for example, on wireless nodes mounted on road infrastructure, unmanned aerial vehicles, etc.), preferably with fixed point architecture. On such platforms, resource utilization, execution time, power consumption, and reconfigurability become critical design parameters that must be optimized. The mapping of algorithms on such platforms is essential to validate their feasibility on the hardware, and the availability of hardware IPs is an essential step toward commercialization. In this work, we focus specifically on the digital hardware implementation of the RSP component of the 802.11ad-based ISAC receiver on the multi-processor system on chip (MPSoC) and consider multiple mobile target detection and localization. The objective of the work is to provide the research community with RSP framework and reconfigurable hardware IPs obtained via mapping of the RSP algorithm on the MPSoC using hardware-software co-design (HSCD), and word-length (WL) analysis. We perform in-depth performance analysis and demonstrate functionality via Graphical User Interface (GUI).

The field programmable gate array (FPGA) based MPSoC is one of the popular edge platforms that offers scalable architecture with complete flexibility and upgradability at software, interface, and hardware levels. One such platform is Zynq MPSoC from AMD-Xilinx, comprising quad-core ARM processor as processing system (PS), i.e., software, and Ultrascale FPGA as programmable logic (PL), i.e., hardware\footnote{In this manuscript, PL, FPGA, and hardware are used interchangeably from hereon. Similarly, PS, ARM processor, and software are used interchangeably.} \cite{ZynqMPSoC1}. From a demonstration and analysis perspective, the MPSoC platform enables the mapping of algorithms entirely on PS or PL or partitioning between PS and PL via HSCD.  In \cite{drozdenko2017hardware}, HSCD of wireless transceivers for WLAN 802.11a protocol was developed on Zynq SoC; while \cite{agrawal2019performance} explored the HSCD of orthogonal frequency division modulation (OFDM) based $L$-band digital aeronautical communication system transceivers. In \cite{schweizer2021fairy} OFDM-based pulse-modulated continuous wave radar prototype was implemented on Zynq MPSoC. In \cite{zhong2016design}, the scattered radar signal corresponding to AC1O automotive radar \cite{ac10_radar} was processed on a Zynq SoC for localization of the target in terms of range and azimuth. The work presented in \cite{Rohit_WSDL} explores deep learning-based wideband spectrum sensing on Zynq SoC. 

The work presented in this paper deals with the design and implementation of three-dimensional RSP for an 802.11ad-based ISAC on the Zynq MPSoC platform. Our detailed literature review shows limited works have been done in this direction. For easier understanding, we compare the existing works with the proposed work in terms of important design features and signal processing objectives in Table.\ref{tab:literature_review}. The main contributions are summarized as follows:
\begin{itemize}
    \item {We model the complete base station (BS) transmitter with ISAC waveform, wireless channels, radar target models, BS receiver with RSP and PHY, and mobile user (MU) PHY on the PS platform of the Zynq MPSoC using the PYNQ framework.}
    \item We develop a novel RSP framework and hardware IPs to estimate the range, azimuth, and Doppler velocity of multiple MU targets. The hardware IPs are optimized via WL optimization, serial-parallel architecture, and HSCD. 
    \item We validate the functional correctness of each IP for a wide range of signal-to-noise ratios (SNR), multiple targets, and line of sight (LoS) and Rician wireless channels in terms of radar detection and localization metrics. We also demonstrate the gain in execution time due to FPGA-based acceleration.
    \item We integrate the complete RSP on Zynq MPSoC via HSCD and explore various configurations to lower execution time for given power and resource constraints. 
    \item We demonstrate the on-the-fly reconfigurable nature of the proposed architecture to support different number of packets, azimuth, and Doppler velocity resolutions. 
    \item { We integrate the RSP accelerator with wireless PHY for BS and MU on Zynq MPSoC and demonstrate the gain in bit-error-rate (BER) and throughput over IEEE 802.11ad.}
\end{itemize}
Our paper is organized in the following manner. {The following section discusses the ISAC signal model for radar and communication along with the RSP framework.
In Section \ref{Sec:RSPArch}, we present the hardware architecture for the RSP and the ISAC system integration on MPSoC. This is followed by a detailed RSP performance analysis in \ref{Sec:PR} and complexity analysis in Section~\ref{Sec:CC}. In Section~\ref{Sec:ReconfA}, we discuss the reconfigurable RSP architecture, followed by Section \ref{Sec:FFTvsMUSIC}, where we compare two Doppler velocity estimation architectures. In \ref{Sec:comm_performance}, we present ISAC performance analysis, and Section \ref{sec:Conclusion} concludes the paper.}


\emph{Notation:} Vectors and matrices are denoted with boldface lower and upper case characters, respectively, while variables are denoted with regular characters.  Vector superscript $T$ denotes the transpose operation, while the symbol $\ast$ denotes the complex conjugate transpose operation.  Time domain and frequency domain representations of a vector are $\mathbf{x}$ and $\tilde{\mathbf{x}}$ respectively. We use the square braces, $[\cdot]$, to indicate discrete-time sequences and the curly braces, $(\cdot)$, to indicate continuous time signals. The symbols $\mathbb{B}$, $\mathbb{R}$, and $\mathbb{C}$ are used to represent binary, real, and complex data, respectively.
\section{ISAC System and Signal Processing Framework}
\label{Sec:SM}
 
\begin{figure}[!b]
 \vspace{-0.5cm}
    \centering
    \includegraphics[scale = 0.365]{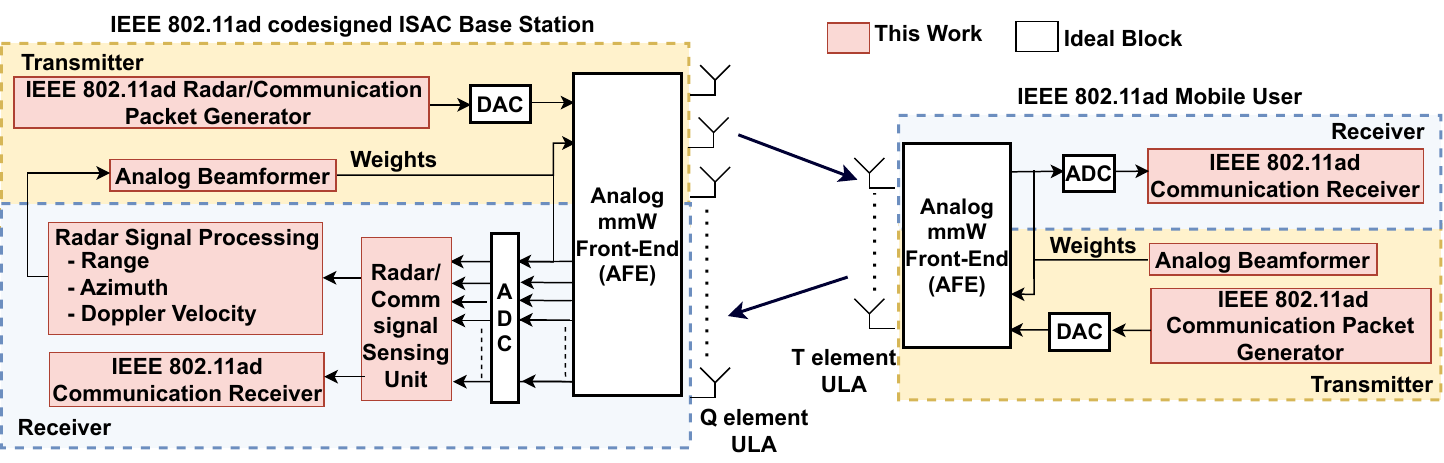}
    \vspace{-0.25cm}
    \caption{\footnotesize Architecture overview of the IEEE 802.11ad ISAC system.}
     \vspace{-0.2cm}
    \label{fig:jrc_arch}
\end{figure}
{In this section, we discuss the radar and communication signal models and corresponding frameworks for the proposed ISAC system. We briefly recapitulate the ISAC architectural framework originally presented in \cite{sneh2022ieee} and then detail the RSP functionality that will be mapped into the novel hardware accelerator. Fig.\ref{fig:jrc_arch} shows an IEEE 802.11ad-based ISAC BS and MU equipped with an 802.11ad-based communication transceiver. The primary goal of the RSP unit in this architecture is the rapid and accurate detection and localization of MUs so that a high-gain communication service can be initiated between the BS and MU as early as possible. The ISAC processing is divided into two segments. The first phase, termed \emph{Stage-1}, is the beam alignment process between the BS and MU. The second, termed \emph{Stage-2}, is the communication phase, where both BS and MU are aligned with each other and communicate over high-directional narrow beams. The beam alignment at BS towards MU is supported by RSP in the proposed ISAC architecture, in contrast to the lengthy beam-search procedure employed in the standard 802.11ad architecture. This offers a shorter stage-1, thereby enabling more time for stage-2 and leading to better link throughput performance over the standard architecture.}

{The IEEE 802.11ad frame structure includes the preamble, header, data, and optional beam refinement (BRF) fields. During stage-1, the ISAC BS transmits a train of $N$ short radar packets, denoted as ${s}_{{tx}_n}$, ($n = 1 \cdots N$), each comprising of Golay sequence (section of the preamble field of 802.11ad frame). 
This is immediately followed by the transmission of $T$ communication downlink packets, ${U}_{{DL}_t}$, ($t = 1 \cdots T$) each with preamble, header, and data fields. These radar and communication pulse trains are meant for beam alignment at the BS and MU, respectively, and are generated in the 802.11ad ISAC packet generator of BS transmitter, which supports single carrier (SC) modulation for the preamble field and orthogonal frequency division multiplexing (OFDM) modulation for the header and data fields as described in \cite{noauthor_ieee_2016-1}. Then, they are passed through the analog-front-end (AFE) and radiated with an omnidirectional beam from a single antenna of the antenna array. The radar signal, ${s}_{tx}$, is scattered from multiple mobile users/targets and arrives at the BS receiver after two-way propagation. A $Q$ element uniform linear array (ULA) is used in the BS receiver, where each antenna element is connected to a distinct mmW chain (amplifier, demodulator, and analog-to-digital converter). The received signal at the BS is first passed through a sensing unit, which distinguishes between a target scattered radar signal or communication signals (uplink or target scattered downlink signal) via correlation and then forwards it to the respective processing block. The radar signal gathered across multiple receiver channels, is processed through digital beamforming in the RSP block to support the detection and localization of the MUs. The estimated angle-of-arrival (AoA) is passed as feedback to the beamformer unit where the beam alignment weights are generated, and the subsequent transmission and reception take place through analog beamforming during stage-2. The downlink signal, ${U}_{DL}$, arrives at the MU receiver after one-way propagation. The MU is equipped with $T$ antennas, and each packet is received by a different beam (out of $T$ total beams) via analog beamforming. The MU performs correlation for all the received packets and selects the beam producing the highest gain as its best beam. Stage-1 is completed once both BS and MU have determined the most optimum beam towards each other. Since $Q>T$, we assume the beam alignment process to complete much faster for MU as compared to BS, and hence, the stage-1 duration is determined by the RSP latency at BS.}

{During stage-2, both MU and BS communicate with each other over the directional link via analog beamforming. The received uplink packets, ${U}_{UL}$, at the BS, and downlink packets, ${U}_{DL}$, at MU are processed by the 802.11ad-based OFDM receiver. Stage-2 continues until the received SNR, either at BS or MU, falls below a certain threshold, in which case stage-1 is restarted. In this work, we treat the digital-to-analog converters (DAC), analog-to-digital converters (ADC), and analog/mmW front end as ideal blocks.
}
\vspace{-0.3cm}
\subsection{Radar Signal Model}
\label{sec:SM}
As noted in prior works \cite{kumari_ieee_2018}, the preamble of each $n^{th}$ IEEE 802.11ad PHY frame consists of $K$-length Golay sequences, $\mathbf{g}_n \in \mathbb{B}^{K \times 1}$, that have perfect autocorrelation properties (zero sidelobes), under idealized conditions to enable range estimation of remote targets. When the order of $N$ Golay sequences is carefully chosen as per the recommendations of \cite{duggal2019micro,duggal2020Doppler}, the resulting two-dimensional signal $\mathbf{G} = [\mathbf{g}_1,\cdots,\mathbf{g}_N]$ is useful for range and Doppler estimation. These sequences constitute the radar signal transmitted from BS during stage-1. Using AFE, the resulting analog transmitted signal is 
\par\noindent\footnotesize
\begin{align}
\mathbf{s}_{tx}(t) = \sum_{k=0}^
{K-1} \mathbf{G}[k,n]\delta(t-kT_s-nT_{PRI}),
\end{align}
\normalsize
 where $T_s$ is the sampling time for each bit in $\mathbf{g}_n$ and $T_{PRI}$ is the interval between two consecutive transmissions. Note that the duty cycle of the waveform $KT_s/T_{PRI}$ is quite low since the Golay sequences in the preamble form a small section within the entire packet. Second, the optional training fields in the packet that are usually included for mmW beam training are omitted since beam selection is being carried out through radar. The duration of the entire signal is one coherent processing interval comprising $NT_{PRI}$. The signal is then suitably convolved with a transmit shaping filter, up-converted to mmW wavelength, $\lambda$, and radiated through the transmitting omnidirectional antenna. Assume that there are $P$ targets within the radar field of view such that each $p^{th}$ target is located at a range $r_p$, azimuth $\phi_p$ and moving with radial velocity $v_p$ with respect to the radar. Then radar echoes from all $P$ targets superpose on the $Q$ element ULA, with $d$ inter-element spacing, of the receiver. Each $q^{th}$ antenna element is supported by a corresponding RF/mmW processing chain comprising a low noise amplifier, in-phase-quadrature demodulator, filter, and analog-to-digital converter. Hence, the received signal at the radar receiver, after down-conversion and digitization, is a three-dimensional data cube across fast time ($k = 1 \cdots K$), slow time ($n = 1 \cdots N$) and antenna elements ($q = 1 \cdots Q$), given by $\mathbf{\mathcal{S}}_{rx} \in \mathbb{C}^{K\times Q \times N} = [\mathbf{S}_{{rx}_1},\cdots \mathbf{S}_{{rx}_N}]$. Here, the radar data square for $n^{th}$ packet, $\mathbf{S}_{{rx}_n} \in \mathbb{C}^{K \times Q}$, is modeled by the generalized Rician channel model as 
 \par\noindent\footnotesize
\begin{align}
\label{eq:RxSig}
\begin{split}
\mathbf{S}_{{rx}_n}= \sum_{{p=1}}^P \left[a_p\sqrt{\frac{J}{J+1}} + \rho\sqrt{\frac{1}{J+1}}\right]\\ \mathbf{G}[n,k-k_p]e^{-j 2\pi f_{D_p}nT_{PRI}}e^{j\frac{2\pi}{\lambda}d(q-1)\sin\phi_p} \\
+ \sum_{i=1}^I\mathbf{\rho_i}\mathbf{G}[n,k-\mathbf{k_i}]e^{-j 2\pi \mathbf{f_{D_i}}nT_{PRI}}e^{j\frac{2\pi}{\lambda}d(q-1)\sin\mathbf{\phi_i}} + \zeta.
\end{split}
\end{align}
\normalsize
\color{black}
Here, $a_p$ is the strength of each $p^{th}$ target's returns which incorporates the target's radar cross-section and the two-way line-of-sight (LOS) path loss factor between the monostatic radar and target; while $k_p$ is the sample index corresponding to the range-induced delay across fast time data; $f_{D_p}$ is the velocity-induced Doppler shift across the slow time data, and $(q-1)d \sin\phi_p$ is the path delay to each $q^{th}$ antenna element. In the above model, we have assumed that the strength and the velocity of the MU are constant within the coherent time interval. The Rician factor $J$ models the contributions of LOS versus the non-LOS (NLOS) in the returns. Thus, the high value of $J$ corresponds to zero multipath, while when $J$ is very low, then the contribution is entirely from multipath with negligible returns through LOS. $\zeta$ in \eqref{eq:RxSig} models complex additive white Gaussian noise. We also incorporate the signals arising from clutter and ghost targets as $I$ discrete scatterers with strength $\rho_i$, range-induced delay $\mathbf{k_i}$, azimuth $\mathbf{\phi_i}$ and Doppler $\mathbf{f_{D_i}}$. Static clutter scatterers are modeled with zero Doppler shift. The strength of these clutter is modeled using an exponential distribution with $\sigma_{i_{avg}}$ average cross-section and total number of these components equals the number of bins in the radar field-of-view. In other words, every bin of the radar data cube consists of returns from one or more scatterers either from target, multipath or clutter along with noise.
\vspace{-0.2cm}
\subsection{Radar Signal Processing (RSP) Framework for ISAC}
\label{Sec:RSPF}
The radar data cube is processed in the following manner. We first process the radar data square, $\mathbf{S}_{{rx}_n}$, across each $n^{th}$ packet to obtain the range-azimuth image. For this, we carry out matched filtering (MF) across the fast-time domain and Fourier-based digital beamforming across the multi-channel antenna data. Due to the computational complexity of carrying out cross-correlation across the fast time domain, we implement the MF through matrix multiplication in the frequency domain and subsequently inverse Fourier transform the results back to the time domain. Then we process the radar detections from this image across $N$ packets to estimate the Doppler velocity as shown in Fig.\ref{fig:rsp_block_diagram}. The steps are as follows:
\begin{figure}[!b]
    \centering
    \vspace{-0.3cm}
    \includegraphics[scale = 0.48]{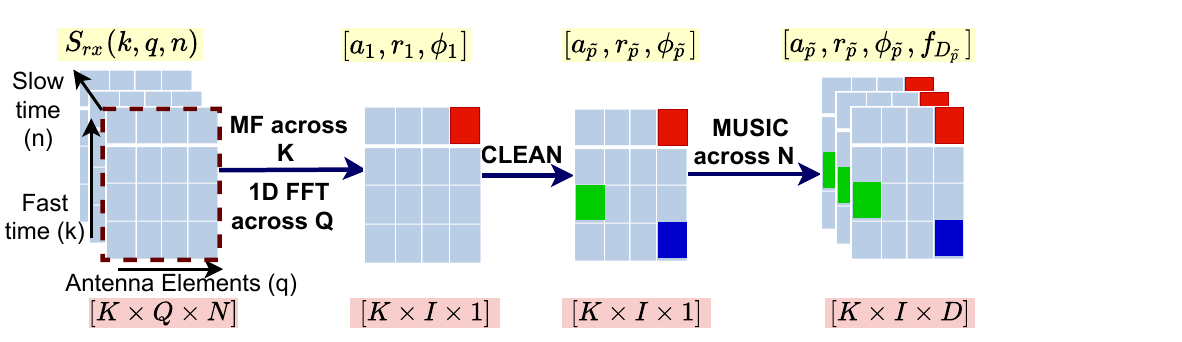}
    \vspace{-0.3cm}
    \caption{\footnotesize RSP of 3-D radar data: MF (in frequency domain) across fast time domain for range, FFT across antennas for azimuth, and MUSIC across slow time samples for Doppler estimation respectively.}
    \label{fig:rsp_block_diagram}
\end{figure}
\textbf{Step.1. One Dimensional Fast Fourier Transform (1D-FFT) on $K$ sample fast time data:} We FFT each $n^{th}$ transmitted Golay sequence to frequency domain ($k_f$):
\par\noindent\footnotesize
\begin{align}
\label{eq:ID-FFT}
\tilde{\mathbf{g}}_n[k_f] = \sum_{k=0}^{K-1} \mathbf{g}_n[k]e^{-j2\pi kT_sk_f},\; k_f = 0 \cdots (K-1)\Delta f,
\end{align}
\normalsize
where $\Delta f = 1/KT_s$ corresponds to the frequency resolution of $k_f$.
\color{black}
Note that this step can be carried out offline (since the transmit sequence is known ahead of real-time operation) in order to reduce real-time computational complexity. Similarly, we FFT the fast time data corresponding to each $q^{th}$ receiver channel from the $n^{th}$ packet: $\tilde{\mathbf{S}}_{{rx}_n}[k_f,q] = \text{1D-FFT}\{\mathbf{S}_{{rx}_n}[k,q]\}$, this time in real-time, similar to \eqref{eq:ID-FFT}.\\
\textbf{Step.2. Digital Beamforming across $Q$ channel data:} We generate $\tilde{\Gamma}_n $ for each $\phi_i, i = 1 \cdots I$ spanning the radar search space through azimuth increments of $\Delta \phi$ and for each $n$ packet, using the operation, 
\par\noindent\footnotesize
\begin{align}
\label{eq:MatMultFFT1}
\tilde{\Gamma}_n[k_f,\phi_i] = \sum_{q=1}^Q\tilde{\mathbf{S}}_{{rx}_n}[k_f,q]\tilde{\mathbf{g}}_n^\ast[k_f] e^{-j k_c d (q-1) \sin\phi_i},\;\; i=1 \cdots I.
\end{align} 
\normalsize
The above operation is carried out for the product of $\tilde{\mathbf{S}}_{{rx}_n}[k_f,q]$ and $\tilde{\mathbf{g}}_n^\ast[k_f]$ in the frequency domain which corresponds to matched filtering in the time domain.\\
\color{black}
\textbf{Step.3. Inverse FFT (IFFT) across $K$ frequency domain data:} Frequency domain data for every $\phi_i$ in $\tilde{\Gamma}_n[\phi_i]$ is converted back to fast time/range ($r_k$) domain via IFFT, 
\par\noindent\footnotesize
\begin{align}
\Gamma_n[r_k,\phi_i] = \sum_{k_f=1}^K \tilde{\Gamma}_n[k_f,\phi_i]e^{\frac{+j4\pi r_k k_f}{c}},\;\; r_k = 0 \cdots \frac{c(K-1)T_s}{2},
\end{align}
\normalsize 
where $c$ represents the speed of light. 
\color{black}
Hence $\Gamma_n$ provides a range-azimuth image of the target space corresponding to every $n^{th}$ slow time sample/packet.\\
\textbf{Step. 4. CLEAN on Range-Azimuth radar images:} Since the radar search space across range and azimuth may consist of several targets, of different scattering strengths, the sidelobes from the point spread response of one target may suppress weaker targets, thereby preventing their detection. Hence, we use \emph{CLEAN} algorithm to detect multiple targets \cite{tsao1988reduction,ram2008through}. First, we initialize a 2D matrix $\chi_1$ to be equal to the range-azimuth space of the first packet ($\Gamma_1$). Then for each $\tilde{p}^{th}$ iteration of the algorithm, we identify the peak amplitude ($\hat{a}_{\tilde{p}}$), and corresponding range ($\hat{r}_{\tilde{p}}$) and azimuth ($\hat{\phi}_{\tilde{p}}$) coordinates of the most dominant target scatterer in the radar search space as shown in equation \eqref{eq:clean1}. Then we obtain the residue, $\chi_{\tilde{p}+1}$, for the subsequent iteration by removing the contributions of $\tilde{p}^{th}$ target from the total response through equation \eqref{eq:clean2}.
\par\noindent\footnotesize
\begin{align}
\label{eq:clean1}
\hat{a}_{\tilde{p}} = \max \chi_{\tilde{p}}, \hat{r}_{\tilde{p}},\hat{\phi}_{\tilde{p}} = \argmax{r_k,\phi_i}\chi_{\tilde{p}-1}
\end{align}
\vspace{-4mm}
\par\noindent\footnotesize
\begin{align}
\label{eq:clean2}
\chi_{\tilde{p}+1}=\chi_{\tilde{p}}-\hat{a}_{\tilde{p}} \mathbf{h}(\hat{r}_{\tilde{p}},\hat{\phi}_{\tilde{p}})
\end{align}\normalsize
In \eqref{eq:clean2}, $\mathbf{h}(\cdot)$ denotes the two-dimensional range-azimuth point spread response of a target located at the specified coordinates within the parentheses. 
The algorithm is terminated when the peak signal detected during an iteration falls below the minimum detectable signal of the radar or when the peak is detected multiple consecutive times at the same location or within the immediate neighborhood of a previously detected target. This results in different numbers of CLEAN iterations depending on the noise and clutter conditions. Second, this algorithm is only applied to the radar data from the first packet, assuming that the targets are present during the first packet. If this assumption is not applicable, then coherent integration of the data from the first few packets must be considered, as discussed in Appendix III. For the subsequent packets, Step 4 is skipped, and we proceed directly to Step.5.\\
\textbf{Step. 5. Multiple signal classification (MUSIC) across slow time:} Finally, once $\tilde{P}$ targets are identified in the range-azimuth space, we estimate their Doppler through the application of MUSIC, a super-resolution algorithm \cite{schmidt1986multiple}, across the slow time domain for every $[\hat{r}_{\tilde{p}},\hat{\phi}_{\tilde{p}}]$ cell as shown in Fig.\ref{fig:rsp_block_diagram}. Essentially, MUSIC is an algorithm that detects signal components by separating the signal from the white noise subspace. It is chosen instead of the more basic FFT since MUSIC requires fewer packets to estimate Doppler. The choice of MUSIC versus FFT is discussed further in Section \ref{Sec:FFTvsMUSIC}. 
First, we vectorize the data across the $N$ slow time packets for each $\tilde{p}^{th}$ target through $ \mathbf{y}_{\tilde{p}} = \left[\Gamma_1[\hat{r}_{\tilde{p}},\hat{\phi}_{\tilde{p}}],\cdots,\Gamma_N[\hat{r}_{\tilde{p}},\hat{\phi}_{\tilde{p}}]\right]$. 
Then, for the target $\tilde{p}$, multiple snapshots are created from a single slow time vector, $\mathbf{y}_{\tilde{p}}$ by splitting it into ${M}=\frac{N}{2}+1$ subarrays, each of length, $S=\frac{N}{2}$ as suggested in \cite{liao2016music}. The averaged covariance matrix, $\mathbf{Y}_{\tilde{P}} \in \mathbb{C}^{S \times S} $,  
is determined by taking the mean of the auto-covariances  generated from each subarray as shown in,
\par\noindent\footnotesize
\begin{align}
\label{eq:music_cov}
\mathbf{Y}_{\tilde{P}} = {\frac{1}{M}}\sum_{m=0}^{M-1} \mathbf{y}_{\tilde{p}}\left[m:m+S-1\right]\mathbf{y}_{\tilde{p}}\left[m:m+S-1\right]^\ast
\end{align}\normalsize
Then, we perform the eigenvector decomposition (EVD) of $\mathbf{Y}_{\tilde{P}} $ by QR-decomposition method \cite{francis1962qr},\par\noindent\footnotesize
\begin{align}
\label{eq:MUSIC1}
 \mathbf{Y}_{\tilde{p}}   = \mathbf{Q}_{\tilde{p}}\mathbf{R}_{\tilde{p}}.
\end{align} \normalsize
Here,  $\mathbf{R}_{\tilde{p}} \in \mathbb{C}^{S \times S} $ converges to an upper triangle matrix and $\mathbf{Q}_{\tilde{p}} \in \mathbb{C}^{S \times S} $  is an orthogonal matrix which comprise of the $\mathbf{\epsilon}_{\tilde{p}} \in \mathbb{C}^{S \times 1}$ and $\mathbf{E}_{\tilde{p}} \in \mathbb{C}^{S \times S-1}$ that correspond to the signal and noise subspaces respectively.
Then, the MUSIC spectrum is generated from \par\noindent\footnotesize
\begin{align}
\label{eq:MUSIC2}
 \mathbf{\mu}_{\tilde{p}}[f_D]  = \frac{1}{\mathbf{\upsilon}^\ast[f_D] \mathbf{E}_{\tilde{p}} {\mathbf{E}}_{\tilde{p}}^\ast  \mathbf{\upsilon}[f_D]}  
\end{align} \normalsize
where, $\mathbf{\upsilon}[f_D]$ is the pseudo-Doppler delay axis for the Doppler frequency $f_D \in \mathbb{R}^{1 \times D}$ spanning from $-f_{D_{max}}$ to $f_{D_{max}}$ with a resolution of $\Delta f_D$.
The Doppler $\hat{f}_{D_{\tilde{p}}}$ corresponding to each $\tilde{p}^{th}$ target is estimated from the coordinate of the peak of $\mathbf{\mu}_{\tilde{p}}$ ($\argmax {f_D} \mathbf{\mu}_{\tilde{p}}$). 
\color{black} Thus, at the end of the exercise, we obtain the estimates of the amplitude ($\hat{a}_{\tilde{p}}$), range ($\hat{r}_{\tilde{p}}$), azimuth ($\hat{\phi}_{\tilde{p}}$) and Doppler frequency ($\hat{f}_{D_{\tilde{p}}}$) of each $\tilde{p}^{th}$ detected target with the precision of $\Delta r, \Delta \phi$ and $\Delta f_D$. The ISAC subsequently uses the Doppler information to distinguish mobile users from static clutter; the range information is used to cluster multiple returns from a single extended target; while the azimuth information is used by the analog beamformer within the ISAC transmitter for establishing directional communication links with the MU.

\emph{Note:} In conventional RSP for standalone radars, range-Doppler processing is performed over every CPI before CLEAN. In our ISAC case, the objective of the RSP is to detect mobile radar targets/users and support communications along their directions. Here, the radar and communication functionalities share the same hardware and spectrum in a time-multiplexed manner. Hence, for a fixed total time interval, the \emph{longer} the radar takes to explore the ISAC field of view and detect and localize the mobile users, the \emph{shorter} is the communication service time available for those mobile users, and lower resultant throughput. This constrains the system to adopt a short radar processing interval to support subsequent communication functionality. The assumption here is that the mobile target is visible during the first pulse of the radar. Hence, the CLEAN algorithm is performed on a two-dimensional range-azimuth data square before Doppler processing.

\subsection{Communication Signal Model for ISAC}
The IEEE 802.11ad downlink packet, comprising $C$ samples, ${u}_{DL}\in \mathbb{C}^{C\times 1}$, is generated at the BS transmitter for reception at the MU. As discussed in \cite{sneh2022ieee}, it includes a preamble field, ${u}_{p}$, used for channel estimation, a header field, ${u}_{h}$, containing control information such as number of data symbols and modulation and coding scheme (MCS) type, and the data fields. As per the standard 802.11ad processing described in \cite{noauthor_ieee_2016-1}, the preamble undergoes SC modulation, and the header and data field are OFDM modulated in the packet generator. The downlink packet is assembled as ${u}_{DL}=[{u}_{p}:{u}_{h}:{u}_{d}]$, at sampling time ${T}_{s}$, and passed through DAC for conversion to analog as shown,
\par\noindent\footnotesize
\begin{align}
    \mathbf{u}_{DL}(t) = \sum_{c=0}^{C-1}\mathbf{u}[cT_s]\delta\left(t-cT_s\right),
\end{align}\normalsize
The analog signal is then upconverted to mmW and transmitted through analog beamforming via Q-element ULA,
\par\noindent\footnotesize
\begin{align}
\mathbf{U}_{DL}(t) = \mathbf{w}_{\phi_p}\mathbf{u}_{DL}^T(t)
\end{align}\normalsize
Here, $\mathbf{w}_{\phi_p}  = [1\; e^{-j k_c d \sin\phi_p} \cdots e^{-j (Q-1)k_c d\sin \phi_p}]^T \in \mathbb{C}^{Q\times 1}$ is the antenna weight vector of the BS ULA, directing the beam along $\phi_p$ in stage-2. During stage-1, there is omnidirectional transmission instead.

The downlink signal travels through the wireless channel and arrives at MU receiver. The received downlink signal at MU after analog beamforming with antenna weight vector, $\mathbf{w}_{\theta}\in\mathbb{C}^{T\times 1}$, and subsequent downconversion and digitization is represented by the Rician channel model as

\footnotesize
\begin{align}
\label{eq:RxSigComm}
\begin{split}
\mathbf{u}_{{rx}}[c]=  \mathbf{w}^T_{\theta}\mathbf{u}_{\theta}\mathbf{u}_{\phi_p}^T\mathbf{U}_{DL}[c-c'_p]\left[a'_p\sqrt{\frac{J}{J+1}} + \rho\sqrt{\frac{1}{J+1}}\right]+\zeta'
\end{split}
\end{align}
\normalsize
Here, $\mathbf{u}_{\phi_p}$ and $\mathbf{u}_{\theta}$ are the steering vectors at BS and MU, respectively; $a'_p$ is the one-way LOS path loss factor between the BS and MU; $c_p'$ is the sample index of the delay induced due to one-way propagation. Rician factor, $J$, models the LOS and NLOS components due to multipath effects, and $\zeta'$ represents the complex additive white Gaussian noise. The digitized signal is then demodulated and decoded at the IEEE 802.11ad OFDM receiver of MU. The 802.11ad uplink packet, ${U}_{UL}$, transmitted by MU for reception at BS, undergoes similar processing as mentioned above.
\color{black}

\section{Reconfigurable RSP Architecture}
\label{Sec:RSPArch}
This section discusses the mapping of the RSP algorithms on the proposed reconfigurable architecture realized on PL. {At the end of the section, we present the complete integrated ISAC architecture with RSP demonstration via a suitable GUI.}
\begin{figure}[!b]
    \centering \includegraphics[width=0.65\linewidth]{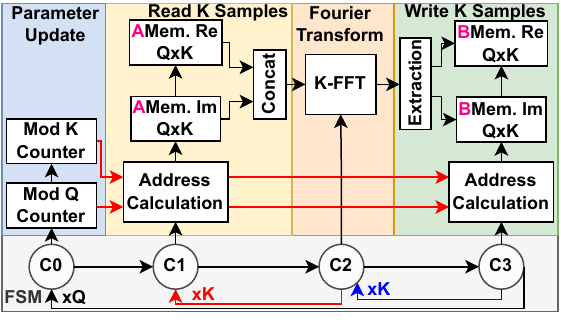}
    \vspace{-0.2cm}
    \caption{\footnotesize One-dimensional FFT on $K$ samples of data from each $q^{th}$ antenna from $n^{th}$ packet.}
    \label{fig:FFT_arch}
     \vspace{-0.2cm}
\end{figure}
\begin{figure}[!b]
\centering
\includegraphics[width=1.02\linewidth]{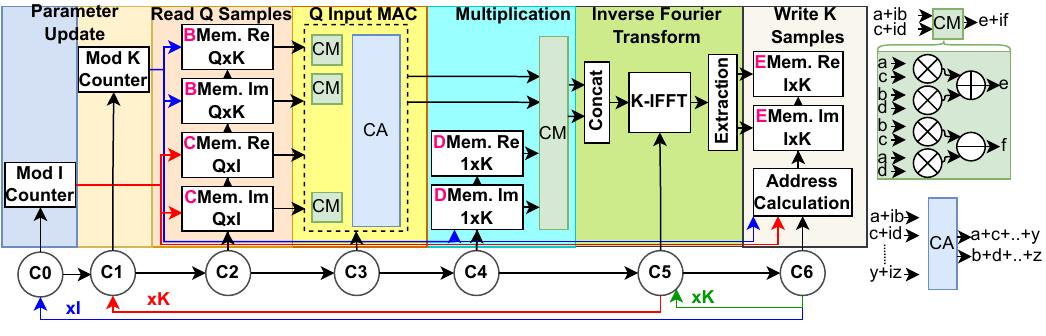}
    \caption{\footnotesize Digital beamforming and Inverse Fourier Transform architecture to estimate range and azimuth of a single target.}
    \label{fig:mf-arch}
\end{figure}
\vspace{-0.6cm}
\subsection{1D-FFT}
The first signal processing task in the proposed RSP architecture is to perform the 1D-FFT on the $K$ samples received from each of the $Q$ antennas, $\mathbf{S}_{{rx}_n} \in \mathbb{C}^{K \times Q}$, (\textbf{Step.1.} in the Section~\ref{Sec:RSPF}). This operation will have to take place in real-time since it operates upon the received radar data. As shown in Fig.~\ref{fig:FFT_arch}, the complex IQ demodulated samples received from each RF/mmW chain are buffered in the block RAM (BRAM) of the FPGA (A) using the direct memory access (DMA). We use separate BRAM for real and imaginary components of the received data. Then, the FFT of size $K$ is performed using the IP provided by AMD-Xilinx, and the output is stored in another BRAM (B). The data movements between FFT IP, BRAMs A and B are controlled using the finite state machine (FSM) depicted in Fig.~\ref{fig:FFT_arch}. Alternatively, we could simultaneously perform the FFT across multiple antennas by using multiple IPs in a serial-parallel fashion to reduce execution time. Since each BRAM has a limited number of read and write ports, we would also need to partition the BRAMs in this strategy, so that the data for multiple FFTs can be read and written simultaneously. 
\vspace{-0.2cm}
\subsection{Digital Beamforming and Inverse FFT}
After 1D-FFT on a single received packet, we have $Q \times K$ complex samples in BRAM B. The digital beamforming across $Q$ antennas, as discussed in \textbf{Step.2.} of RSP, is given as,
\par\noindent\footnotesize
\begin{align}
\label{eq:MatMultFFT}
\tilde{\Gamma}_n[\phi_i] = \text{diag}\left[\tilde{\mathbf{S}}_{{rx}_n}\mathbf{w}_{\phi_i}\tilde{\mathbf{g}}_n^\ast \right]
\end{align} \normalsize
Here, $\mathbf{w}_{\phi_i}  = [1\; e^{-j k_c d \sin\phi_i} \cdots e^{-j (Q-1)k_c d\sin \phi_i}]^T$ provides a digital weight vector to the ULA receiver data to scan the beam at $\phi_i$; and $\text{diag}$ extracts the diagonal elements of the complex $K \times K$ matrix generated after matrix multiplication. \\ The direct hardware realization of equation \eqref{eq:MatMultFFT} is computationally complex and involves multiple ($I$) large-size matrix multiplication operations. Hence, we exploit the significant redundancy associated with (1) the selection of diagonal elements to obtain $\Gamma_n \in \mathbb{C}^{K \times I}$; and (2) the $Q$ replicas of the $\mathbf{g}_n$, to realize a computationally efficient architecture implementation of equation \eqref{eq:MatMultFFT} followed by the corresponding IFFT in \textbf{Step.3.} as shown in Fig.~\ref{fig:mf-arch}.

The matrix, $\mathbf{W} \in \mathbb{C}^{I \times Q}$, comprising of digital weight vector for all $Q$ antennas and $I$ azimuth values, is fixed. Hence, it is precomputed and initialized in BRAM C. In real-time, the first step is to perform element-wise complex multiplications (CM) between each column of BRAM B containing 1D-FFT output and the $i^{th}$ column of BRAM C. These operations are performed using parallel $Q$ CMs. Resultant $Q$ outputs are accumulated using $Q$-input complex adder (CA) to obtain a single complex sample, which is multiplied with the corresponding sample of the pre-computed frequency domain version of the transmitted Golay sequence ($\mathbf{\tilde{g}}_n[k_f], k_f = 1 \cdots K$) stored in BRAM D.  This process is repeated for each $k^{th}$ column of BRAM B, as shown in the figure, to obtain $K$ complex samples for a given azimuth, $\phi_i$. Next, we perform the 1D-IFFT on $K$ samples using AMD-Xilinx IP. This process is repeated for $I$ azimuth values to obtain $I \times K$ samples in BRAM E.

The direct mapping of equation \eqref{eq:MatMultFFT} incurs $QK$ CMs for digital beamforming followed by $K^2Q$ CM and $K^2$ $(Q-1)$-input CA. The proposed computationally efficient architecture also needs $QK$ CMs for digital beamforming. However, it needs only $KQ$ CMs and $K$ $(Q-1)$-input CA. Thus, the proposed architecture needs $K$ times fewer CMs and CAs, offering significant savings in resource utilization, execution time, and power consumption. The BRAM memory requirement is also reduced by a factor $Q$ due to the shifting of CA operations before the element-wise multiplication with the frequency domain version of the transmitted Golay sequence. 

The proposed low-complexity architecture allows further parallelization to improve the execution time. For instance, the digital beamforming and 1D-IFFT are done sequentially $I$ times, one for each azimuth value. Since the computations for any two azimuth values are independent, we have explored the serial-parallel configurations of the architecture in Fig.~\ref{fig:mf-arch}. Specifically, we have used multiple units of CMs, CAs, and IFFT along with the BRAM partitions to allow simultaneous read and write operations. The FSM takes care of appropriate scheduling and movement of data across the parallel blocks. Please refer to Section~\ref{Sec:CC} for more details about the performance gain using our architecture. 
\vspace{-0.2cm}
\subsection{Range and Azimuth Estimation}
The peak search algorithm in \textbf{Step.4. CLEAN operation} then uses the 1D-IFFT output to estimate the amplitude, range, and azimuth of the detected target. Since parallelization of this operation would demand complete partitions of memory, we implement a sequential execution of this operation in the PS section of the MPSoC. To summarize, whenever the new packet is received, the architectures in Fig.~\ref{fig:FFT_arch} and Fig.~\ref{fig:mf-arch} are used by the schedular in the PS to perform the data-intensive FFT, beamforming, and IFFT operations in the FPGA. We refer to this entire operation as MF. The PS reads the data from BRAM E via DMA and performs the peak search. 
\vspace{-0.2cm}

\begin{figure}[!b]
    \centering
     \vspace{-0.2cm}
     \includegraphics[scale = 0.53]{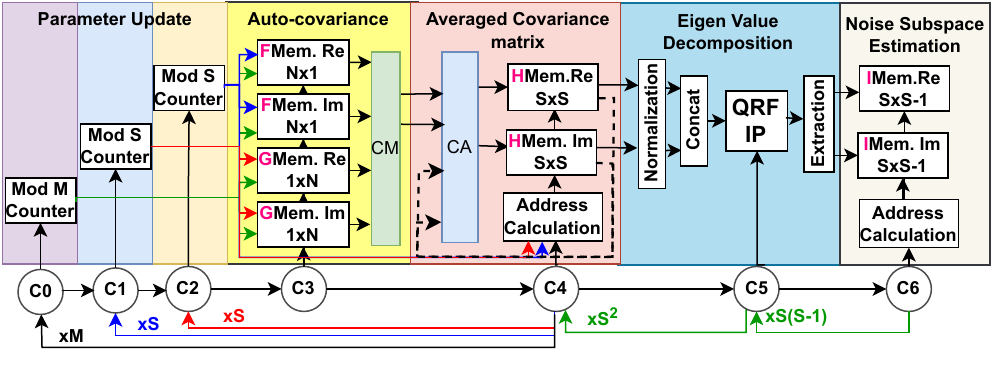}
    \vspace{-0.4cm}
    \caption{\footnotesize Architecture corresponding to covariance matrix generation and EVD stages in MUSIC.}
     \vspace{-0.2cm}
    \label{fig:MUSIC1}
\end{figure}
\begin{figure}[!b]
    \centering
     \vspace{-0.2cm}
     \includegraphics[scale = 0.57]{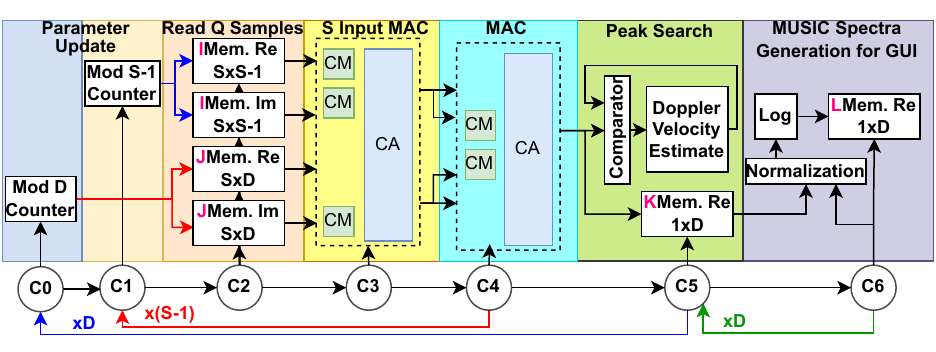}
    \vspace{-0.4cm}
    \caption{\footnotesize Architecture corresponding to MUSIC spectrum generation stage.}
    \label{fig:MUSIC2}
    \vspace{-0.4cm}
\end{figure}
\subsection{Multiple Packets Processing}
The Doppler velocity estimation requires multiple packets received over the antenna sequentially. Each packet is processed by the architectures in Fig.~\ref{fig:FFT_arch} and Fig.~\ref{fig:mf-arch} to obtain the $I \times K$ matrix. The peak selection unit in the PS selects one cell from each $I \times K$ matrix based on the estimated range and azimuth. These $N$ samples extracted from $N$ packets are processed by the MUSIC in Section~\ref{sec:MUSIC} to estimate the Doppler velocity of the detected target. After the first packet, the rest of the packets can be processed in parallel using multiple instances of the architectures in Fig.~\ref{fig:FFT_arch} and Fig.~\ref{fig:mf-arch}.
\vspace{-0.45cm}
\subsection{MUSIC Based Doppler Velocity Estimation}
\label{sec:MUSIC}
The MUSIC algorithm, in \textbf{Step.5.} of the RSP, comprises broadly three tasks: covariance matrix generation, EVD, and MUSIC spectrum generation (MSG). 
The input vector, $\mathbf{y}_{\tilde{p}}$, is formed by selecting the values in the $\tilde{p}^{th}$ target's range-azimuth cell for $n=1:N$ packets after MF. Then, the vector is sent to the MUSIC accelerator in PL from PS through DMA. As shown in Figure~\ref{fig:MUSIC1}, the real and imaginary parts of $\mathbf{y}_{\tilde{p}}$ are stored in BRAM F and that of $\mathbf{{y}_{\tilde{p}}^\ast}$ in BRAM G. With every increment of the MOD $M$ counter, a new subarray of length $S$ is read from BRAMs F and G and is sent to the complex conjugate multiplication unit for auto-covariance operation. The output of the CM unit is simultaneously fed to the CA unit for complex addition with the previously accumulated output stored in BRAM H. In each iteration of the MOD $M$ counter, an $S\times S$ auto-covariance matrix, for the selected input subarray is produced, accumulated, and written in BRAM H. After M iterations, the real and imaginary accumulated sum of the covariances are read from BRAM H, normalized and concatenated to produce the averaged covariance matrix, $\mathbf{Y}_{\tilde{P}}$. As shown in the figure, multiple CM units and the partitioning of the BRAM enable the parallelization of operations in the auto-covariance stage.

Subsequently, EVD is performed on $\mathbf{Y}_{\tilde{p}}$ via QR factorization method with the inbuilt AMD Xilinx QRF IP, wherein the matrix is iteratively decomposed into $\mathbf{Y}_{\tilde{p}}=\mathbf{Q}_{\tilde{p}}\mathbf{R}_{\tilde{p}}$, until $\mathbf{R}_{\tilde{p}}$ converges to an upper triangular matrix. The diagonal elements of the resultant $\mathbf{R}_{\tilde{p}}$ are the eigenvalues arranged in descending order, whereas $\mathbf{Q}_{\tilde{p}}$ is an orthogonal matrix of the respective eigenvectors. $\mathbf{Q}_{\tilde{p}}$ is partitioned into signal and noise subspaces. Since we are computing the Doppler of only a single target (at a specified range and azimuth), the signal subspace, $\mathbf{\epsilon}_{\tilde{p}}$, comprising a single eigenvector corresponding to the largest eigenvalue, is discarded, and the noise subspace $\mathbf{E}_{\tilde{p}}$ with remaining $(S-1)$ eigenvectors is further processed.
\begin{figure*}[!b]
    \centering
    \includegraphics[width=0.85 \linewidth]{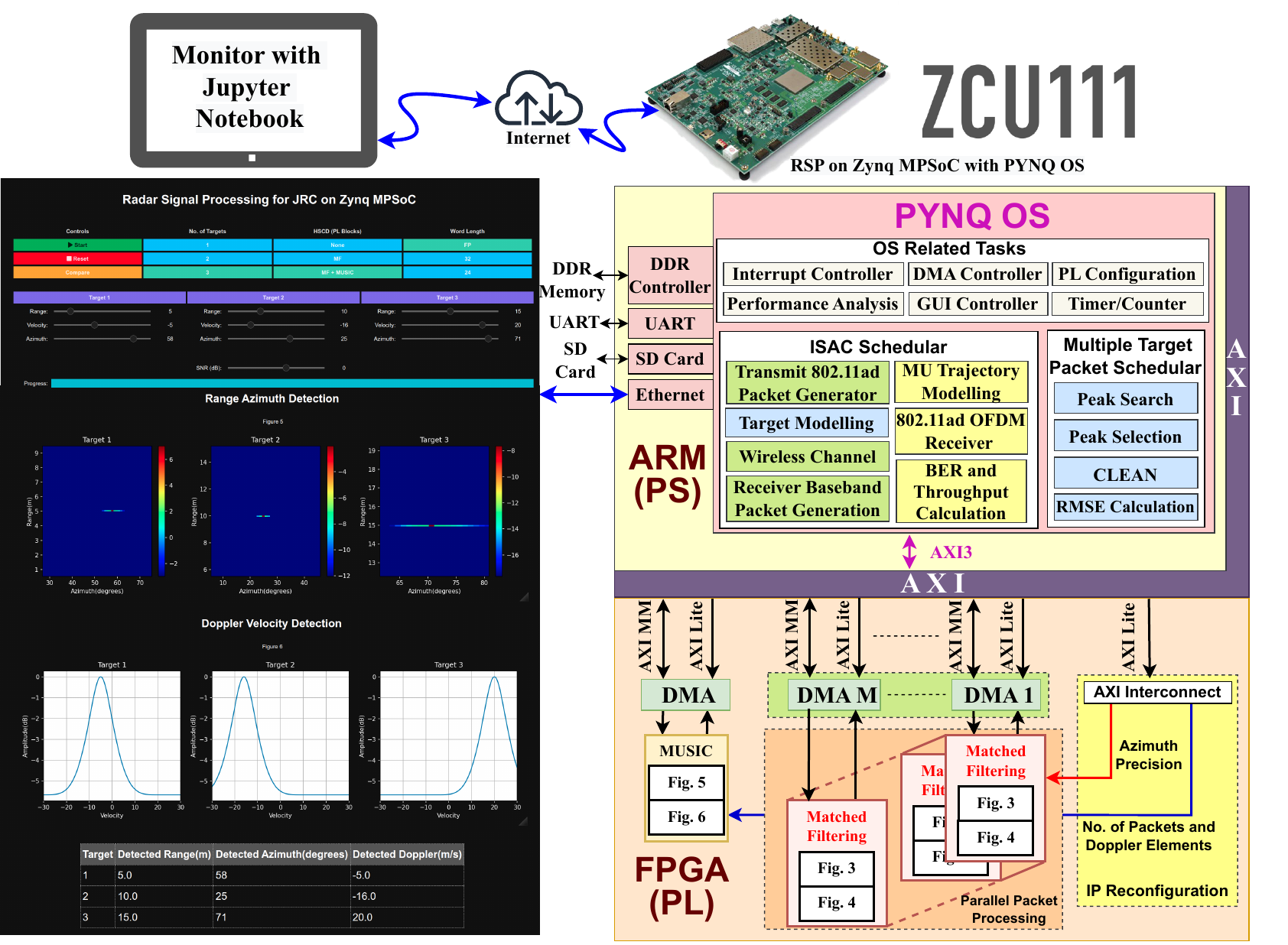}
    \caption{\footnotesize Architecture overview of ISAC implementation on Zynq MPSoC along with GUI. Here, Zynq MPSoC is connected to the institute network and GUI is remotely accessible on any desktop or mobile phone with internet connectivity.}
    \label{fig:comp_arch_GUI}
\end{figure*}
Fig.~\ref{fig:MUSIC2} illustrates the MSG as per equation \eqref{eq:MUSIC2}. The Doppler matrix, $\mathbf{f_D} \in \mathbb{C}^{S \times D}$, is pre-computed and stored in BRAM $J$. In the beginning, element-wise multiplication and accumulation (MAC) are done for a given Doppler element and $(S-1)$ eigenvectors. The output of MAC is squared and accumulated. As shown in the figure, $S$ MAC operations for a given eigenvector are done in parallel but the computations for each $S-1$ eigenvector are done sequentially. This process is repeated for $D$ Doppler elements to obtain the $1 \times D$ size MUSIC spectrum. The peak search unit finds the maximum value to obtain the Doppler velocity of the detected target. The MUSIC spectrum is further normalized and converted to decibels for visualization.
\subsection{RSP Architecture for Multiple Target Detection Using Multiple Packets on MPSoC}
The RSP architecture to detect multiple targets and estimate their peak amplitude, range, azimuth, and Doppler velocity on ZCU111 MPSoC from AMD-Xilinx is shown in Fig.~\ref{fig:comp_arch_GUI}. We have deployed PYNQ OS on the quad-core A53 processor in PS which handles hardware and peripheral initialization, communication with hardware drivers, and IPs. In addition, we have deployed some of the RSP tasks on PS. From the RSP perspective, the first task is to generate the transmit radar packet as per the 802.11ad protocol as discussed in Section~\ref{sec:SM}. We have developed a user-friendly GUI, shown on the left side of the same figure, to allow users to select the number of targets, target parameters, SNR, and architecture for RSP. Then we generate the received baseband packets for $Q$ antennas using wireless channel and radar target modeling in PS. The \textit{ISAC Schedular} process handles these tasks, which are realized in the PS.

The next task is to detect the strongest target and estimate its parameters. The \textit{Multiple Target Packet Schedular} processes in PS sequentially and sends all received packets to the MF block in PL via DMA. Depending on the available resources, the proposed architecture allows the parallel processing of multiple packets using parallel MF blocks and multi-channel DMAs, as shown in Fig.  ~\ref{fig:comp_arch_GUI}. 
Using the MF output of the first packet, the peak search process in PS calculates the range and azimuth of the first detected target. The CLEAN algorithm implemented in PS is used to detect multiple mobile targets in the ISAC field of view by utilizing the parameter information obtained from the previously localized target. The detected amplitude, range, and azimuth of the dominant target in each iteration are used to generate the point spread response ($\mathbf{h}(\cdot)$) as discussed in equation \eqref{eq:clean2} for the subsequent target detection. The \textit{Multiple Target Packet Schedular} processes in PS generate a dummy radar data square signal corresponding to the dominant target in the most recent iteration and processes it using the MF in FPGA to generate the point spread response, which is then subtracted from the original response to obtain the residue for the next iteration. This process is followed by Doppler velocity estimation of the targets using MUSIC. 

Based on the MF outputs of all packets, the peak selection block selects the desired samples from the MF output and forms the packet for the MUSIC block in FPGA. This packet is sent to the MUSIC via DMA and the corresponding Doppler velocity is estimated. Then the FPGA transfers the complete MUSIC spectra for displaying on the GUI. For performance analysis, root mean square error (RMSE) is calculated (as discussed in the later sections) and displayed on the GUI.

The proposed architecture is carefully designed to offer reconfiguration capability, which enables the run-time selection of parameters such as the angular precision ($\Delta \phi$) for the azimuth radiation pattern, the number of packets ($N$), and Doppler elements ($D$) utilized for Doppler estimation. As shown in Fig.~\ref{fig:mf-arch}, we can configure the hardware IP to support different $\Delta \phi$ by adjusting the MOD $I$ counter resolution. The PS does this via the AXI Lite port in the proposed architecture. Similarly, the number of packets in a coherent interval and Doppler elements can be easily controlled by controlling the number of times the DMA is configured since there is one DMA transaction per packet for MF IP. The MUSIC IP is configured for a given packet size and Doppler elements via the AXI Lite port in the proposed architecture. The impact of the reconfigurable architecture on execution time and power consumption is discussed later in Section~\ref{Sec:ReconfA}.

\vspace{-0.2cm}
\subsection{Integrated ISAC architecture on MPSoC}
{The complete ISAC system architecture for communication between the BS and MU, as described in Section \ref{Sec:SM}, is realized on Zynq MPSoC. This is shown in Fig.\ref{fig:comp_arch_GUI}. The MU is modeled to travel along a predetermined trajectory for a fixed time duration (as discussed later in Section IV-B) in PS. The 802.11ad packet generator block in PS generates packets continuously during the simulation time, including the radar packets, which are transmitted by BS, and the communication downlink and uplink packets transmitted by BS and MU, respectively. During stage-1, the \textit{ISAC Schedular} initiates the generation of $N$ radar packets and $T$ downlink packets in the packet generator. It selects between the one-way or two-way propagation models according to the communication or radar channel modeling. Also, the particular baseband received packet generation is selected, specific to the case of communication processing at BS, at MU and radar processing. The target modeling for RSP is performed based on the current position of MU in the trajectory. The tasks related to RSP for $N$ radar packets are scheduled by the \textit{ISAC Schedular} and \textit{Multiple Target Packet Schedular} and carried out in a similar manner as discussed previously, with RSP accelerator in PL. The AoA towards MU is determined. The $T$ downlink packets, arriving at the MU receiver, ${U}_{DL}$ are processed through cross-correlation to select the best beam towards BS. This completes stage-1, and the \textit{ISAC Schedular} now initiates stage-2 with directional transmission and reception at BS and MU. During stage-2, the arriving downlink packets at the MU receiver and the uplink packets at the BS receiver are processed in the 802.11ad OFDM receiver of PS. Additionally, the performance in terms of communication link metrics is also evaluated in PS. The \textit{ISAC Schedular} schedules stage-1 and stage-2 by continuously monitoring the received SNR at BS and MU throughout the simulation. The stage-1 restarts once the received SNR falls below a certain threshold. It also supports an adaptive modulation scheme, where it can switch between high-order modulation for high SNR and low-order for lower SNR for better link throughput.
}
\section{RSP Performance Analysis}
\label{Sec:PR}
In this section, we validate the functional correctness of the proposed RSP hardware architecture for different WL, SNR, and number of targets. We use root mean square error (RMSE) between the estimated range/azimuth/Doppler (denoted as $\hat{x}$) and their ground truth counterpart values (denoted as $x$) as the performance metric, given as, \small $RMSE = \sqrt{\frac{\sum_{z=1}^{Z} (x - \hat{x}) ^ 2}{Z}}$\normalsize.
\subsection{Simulation Model}
\begin{figure}[!ht]
    \centering
    \includegraphics[scale = 0.23]{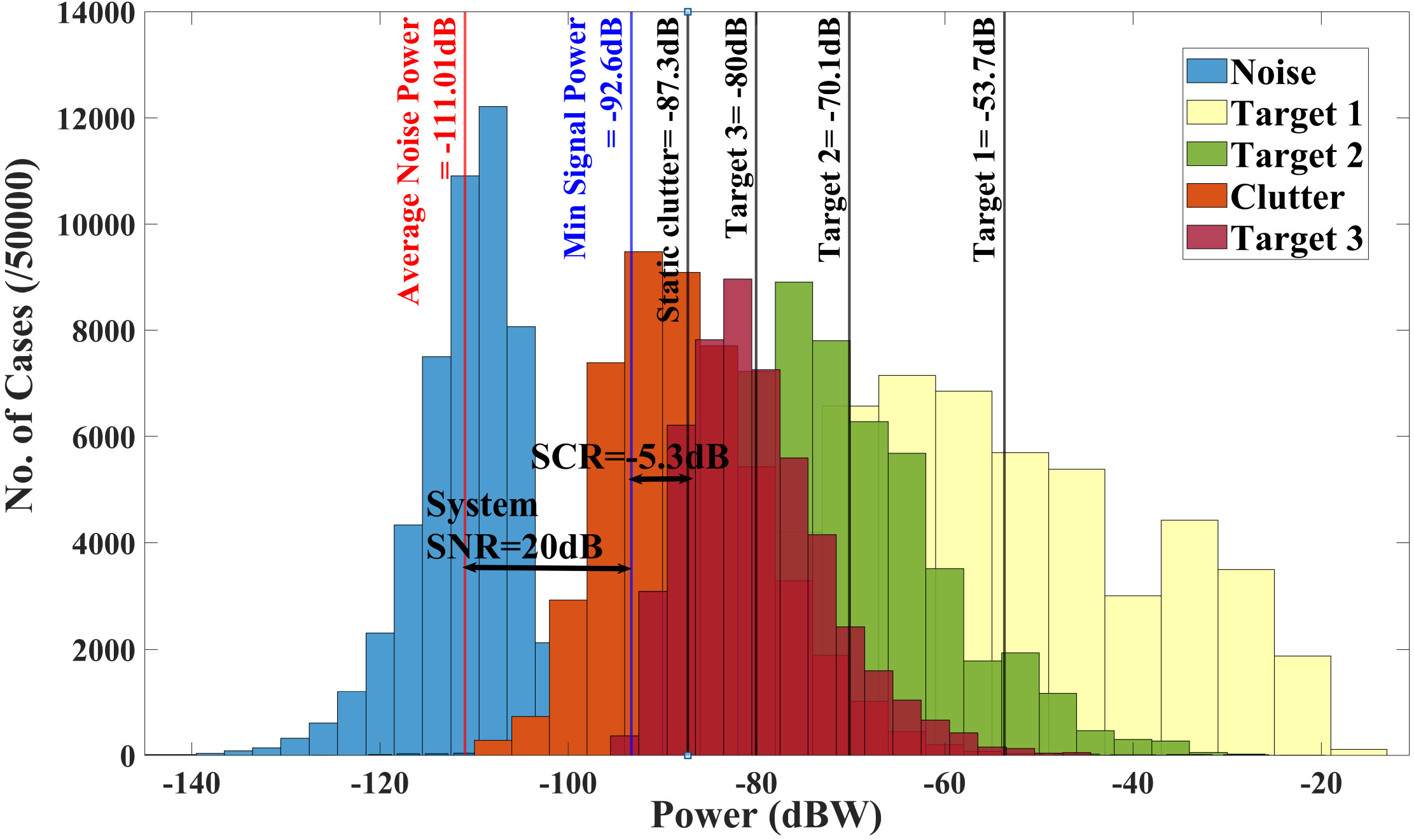}
    \vspace{-0.3cm}
    \caption{\footnotesize Distribution of the received signal strengths of the three targets, clutter scatterers, and noise.}
    \label{fig:static_clutter_hist}
    \vspace{-0.3cm}
\end{figure}

In our work, we have considered a total of $Q=32$ antennas and $N=100$ packets/slow time samples, each consisting of $K=1024$ fast time samples generated by zero padding a Doppler-resilient Golay sequence of 512 bits. 
The radar field-of-view spans from 0 to 40 m along the range with $\Delta r = 0.085$ m; azimuth spans from $-90^{\circ}$ to $90^{\circ}$ at $\Delta \phi = 1^{\circ}$; and the Doppler velocity spans $-30$ m/s to $+30$ m/s with a velocity resolution of 0.3 m/s. All the RMSE results presented hereafter are obtained after averaging over 200 different experiments. 
In each experiment, we consider a three-target scenario where each target is treated as a point target whose position is a random vector uniformly distributed in the radar's field of view. The radar cross-section of each target is also a random variable and follows the Swerling-1 (exponential) probability distribution, which is most common for radar targets \cite{skolnik1983introduction}. The average RCS of the first, second, and third targets are 10, 5, and 3 square meters, respectively, resulting in the distribution of the returned signals from these targets as shown in yellow (T1), green (T2), and maroon (T3) in Fig.\ref{fig:static_clutter_hist}. Further, we model noise voltage as a complex Gaussian variable with mean zero, standard deviation of $\sqrt{N_p}$, and random phase sampled from a uniform distribution between 0 and $2\pi$. The resulting noise power distribution is shown in blue. We define the system SNR as the ratio between the minimum signal that can be detected by the radar receiver, $S_{min}$, and the mean noise power of the receiver electronics. $S_{min}$, as indicated in the figure, is computed based on the average cross-section of the weakest radar target (T3) at the greatest possible range from the radar. Hence, the SNR is a system metric and is not identical for T1, T2, and T3. In fact, T1, T2, and T3 have mean SNRs that are offset by 39 dB, 22.5 dB, and 12.6 dB from the system SNR as shown in the figure. This implies that when the system SNR is 0 dB, the mean SNR for T1, T2, and T3 is 39 dB, 22.5 dB, and 12.6 dB, respectively. We consider a wide range of system SNR from -15 dB to 20 dB to verify the functional correctness of fixed-point hardware IPs. In practice, the SNR range is expected to be 0 dB or higher.\\
In clutter scenarios, we consider the average RCS of the discrete clutter scatterers to be 0.2$ m^2$ giving rise to the orange distribution with mean clutter power of -87.3 dB as shown in Fig.\ref{fig:static_clutter_hist}. Similarly, the signal-to-clutter ratio (SCR) is also a system metric and is defined as the ratio between the minimum signal that can be detected by the radar receiver, $S_{min}$, and the mean clutter power. The effect of static clutter on radar performance is discussed in detail in Appendix II.
\begin{figure*}[!b]
    \centering
    \includegraphics[scale = 0.47]{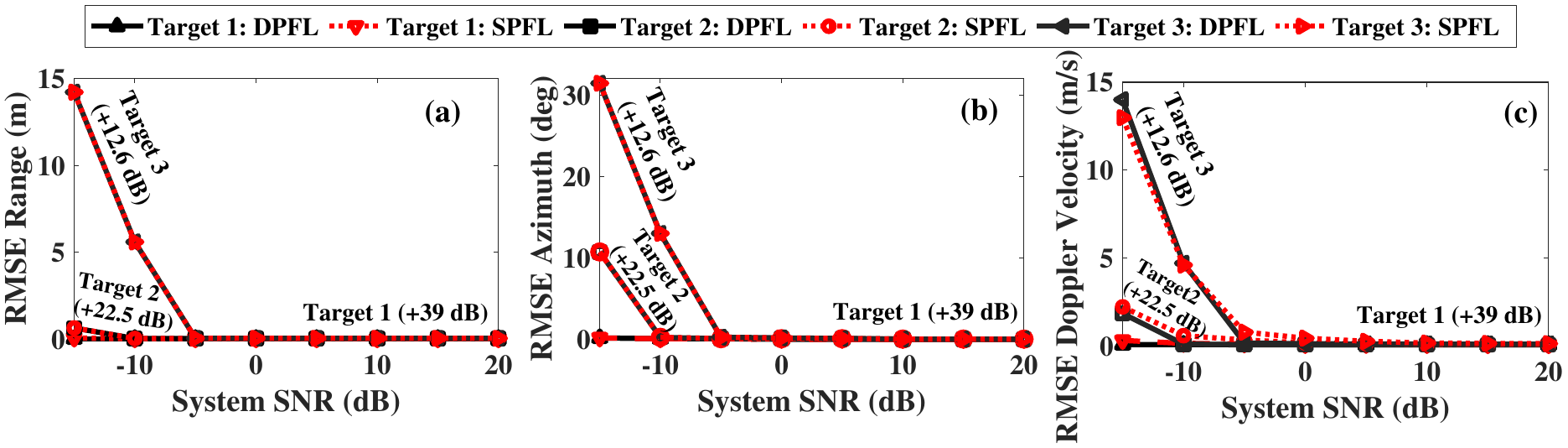}
    \vspace{-0.3cm}
    \caption{\footnotesize Functionality verification of floating point architectures for three targets in terms of RMSE of range, azimuth, and Doppler velocity estimations with LOS channel. Based on strength, targets 1, 2, and 3 have offsets of 39 dB, 22.5 dB, and 12.6 dB, respectively, with respect to the system SNR.}
    \label{fig:rmse_FL}
\end{figure*}

\subsection{Floating Point Architecture}
\label{Sec:SPFL}

\textbf{Radar Detection Metrics:} In any particular simulation experiment, the targets' strengths - determined through the radar range equation - and positions can vary considerably. Therefore, any two targets may have similar strength at the radar receiver or a weaker target may lie within the main lobe of the stronger target. Under these conditions, there may be instances when the weaker target may remain undetected and the stronger target may be detected multiple times because the complete response of the stronger target is not removed by the CLEAN operation. These are counted as false alarms (FA). Further, noise samples that are above the threshold may also give rise to FA. We present the missed detections (MD) and FA for each of the three targets obtained from the 200 simulations in Table.\ref{tab:MD_FL_Rician}.

The results also show that the number of FA and MD are lowest for the strongest target (target 1) followed by the second and third targets for all the system SNR values. Due to the imperfect removal of the point spread response of the strongest target, the second and third targets are misidentified in the locations of the first in subsequent trials in some cases. As the SNR deteriorates, the number of FA due to noise increases.
\color{black}
\begin{figure*}[!ht]
    \centering
    \includegraphics[scale = 0.47]{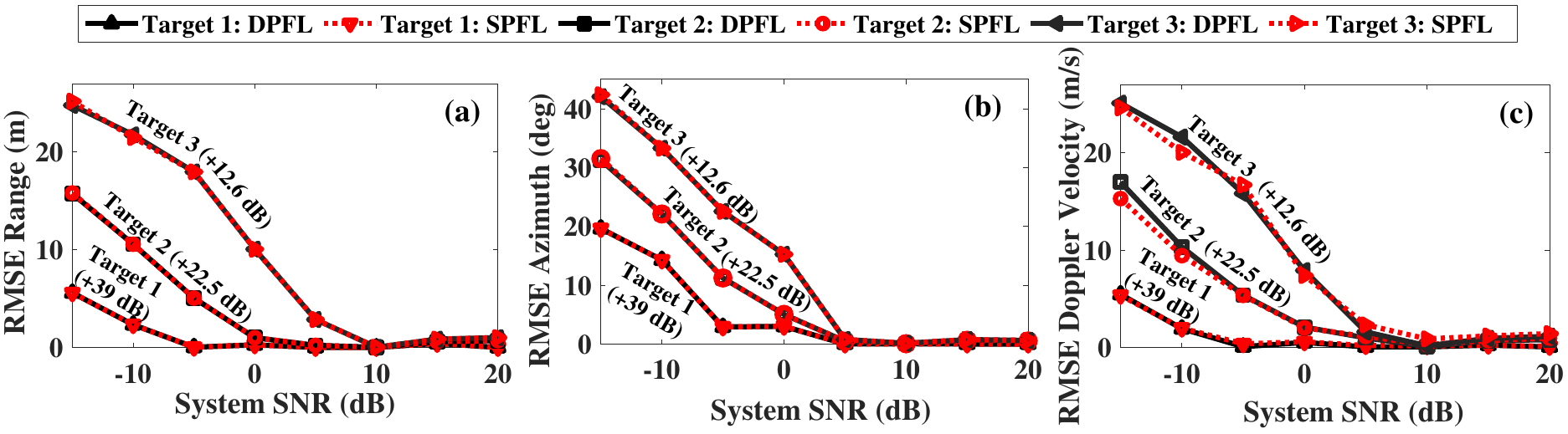}
    \vspace{-0.2cm}
    \caption{\footnotesize Functionality verification of floating point architectures of RSP in terms of range, azimuth, and Doppler velocity RMSE for three targets with Rician channel.}
    \label{fig:rmse_rician_FL1}
\vspace{-0.3cm}
\end{figure*}

\textbf{Radar Localization Metrics:} In Fig.~\ref{fig:rmse_FL}(a), we benchmark the range detection performance of the single precision floating point (SPFL) WL on the MPSoC with the double precision floating point (DPFL) WL in PS or MATLAB for LOS conditions. 
The RMSE of both the MPSoC and MATLAB approaches is nearly identical for all targets and system SNRs, validating the functional correctness of the proposed hardware architectures for range estimation. The results show that the RMSE is well below 0.5\% for the first (or strongest) target for all SNRs. The RMSE degrades for the second and third targets. However, even here, the RMSE decreases with the increase in SNR as expected. The poorer performance for the weaker targets is due to multiple reasons: First, the third detected target is likely to have a lower peak amplitude compared to the second detected target which will have a lower amplitude than the first target due to the nature of the algorithm; Second, the weaker targets may lie within the main lobe of the stronger targets; Finally, due to the superposition of the returns from multiple targets, the CLEAN algorithm can never fully remove the contributions of the dominant scatterer in equation \eqref{eq:clean2}. This will result in error propagation across the iterations of the algorithm. The degradation becomes significant at low SNR. Numerically, the RMSE is lower than 0.5\% for SNR higher than -5dB and 5 dB for the second and third targets, respectively. In Fig.~\ref{fig:rmse_FL}(b), we compare the azimuth detection performance of both approaches. We observe similar behavior for SNR and multiple targets. In both cases, we have used a single packet to estimate the range and azimuth of all targets. 
Next, we compare the RMSE in Doppler velocity estimation in Fig.~\ref{fig:rmse_FL}(c). There is no difference between the two approaches validating the functional correctness of the complete RSP architecture, including iterative CLEAN algorithm, peak search, point spread response generation, and MUSIC realized on the MPSoC. 

We repeat the LOS experiments with Rician channel conditions where $J=7dB$  which corresponds to a rural scenario \cite{zhu2014probability} and DPFL framework and report the radar detection metric results in Table.\ref{tab:MD_FL_Rician}.
\renewcommand{\arraystretch}{1.2}
\begin{table}[!ht]
\footnotesize
\caption{\footnotesize Radar Detection metrics: False alarms (FA) and Missed Detections (MD) for LOS and Rician channel}
\centering
\resizebox{0.4\textwidth}{!}{%
\begin{tabular}{|c|cc|cc|cc|}
\hline
\multirow{3}{*}{\begin{tabular}[c]{@{}c@{}}\textbf{System}\\ \textbf{SNR}\end{tabular}} &
  \multicolumn{2}{c|}{\textbf{Target 1 (/200)}} &
  \multicolumn{2}{c|}{\textbf{Target 2 (/200)}} &
  \multicolumn{2}{c|}{\textbf{Target 3 (/200)}} \\ \cline{2-7} 
             & \multicolumn{2}{c|}{\textbf{(FA, MD)}} & \multicolumn{2}{c|}{\textbf{(FA, MD)}} & \multicolumn{2}{c|}{\textbf{(FA, MD)}}  \\ \cline{2-7} 
 &
  \multicolumn{1}{l|}{\textbf{LOS}} &
  \multicolumn{1}{l|}{\textbf{Rician}} &
  \multicolumn{1}{l|}{\textbf{LOS}} &
  \multicolumn{1}{l|}{\textbf{Rician}} &
  \multicolumn{1}{l|}{\textbf{LOS}} &
  \multicolumn{1}{l|}{\textbf{Rician}} \\ \hline
\textbf{20}  & \multicolumn{1}{c|}{(0,0)}   & (0,0)   & \multicolumn{1}{c|}{(8,0)}   & (0,3)   & \multicolumn{1}{c|}{(29,0)}  & (3,27)   \\ \hline
\textbf{10}  & \multicolumn{1}{c|}{(0,0)}   & (0,1)   & \multicolumn{1}{c|}{(7,0)}   & (0,6)   & \multicolumn{1}{c|}{(30,0)}  & (3,28)   \\ \hline
\textbf{0}   & \multicolumn{1}{c|}{(0,1)}   & (0,1)   & \multicolumn{1}{c|}{(7,5)}   & (0,19)  & \multicolumn{1}{c|}{(30,3)}  & (4,76)   \\ \hline
\textbf{-10} & \multicolumn{1}{c|}{(0,1)}   & (0,16)  & \multicolumn{1}{c|}{(3,5)}   & (1,55)  & \multicolumn{1}{c|}{(31,20)} & (16,133) \\ \hline
\end{tabular}
}%
\label{tab:MD_FL_Rician}
\end{table}
The results show that the radar detection metrics deteriorate for the Rician conditions compared to the LOS conditions. The most significant deterioration is observed for the weaker targets compared to the stronger targets as the number of MD increases due to the NLOS component. Next, we consider the radar localization metrics presented in
Fig.\ref{fig:rmse_rician_FL1}. 
The results show that the RMSE of the targets is significantly higher for Rician conditions compared with the results for LOS conditions shown in Fig.\ref{fig:rmse_FL}. Again, the weakest/third target shows the highest RMSE followed by the second target and then the first. The performance of
 the algorithms can be improved through coherent integration
 of the data across multiple time snapshots. This is discussed
 further in Appendix III.

 \begin{table*}[!b]
 \footnotesize
\centering
\caption{\footnotesize Radar detection metrics: False alarms (FA) and Missed Detections (MD) for different fixed point architectures}
\label{tab:FA_MD_WL}
\renewcommand{\arraystretch}{1.2}
\resizebox{0.8\textwidth}{!}{%
{
\footnotesize
\begin{tabular}{|c|cccc|cccc|cccc|}
\hline
\multirow{3}{*}{{\textbf{\begin{tabular}[c]{@{}c@{}}\textbf{System}\\ \textbf{SNR}\end{tabular}}}} & \multicolumn{4}{c|}{\textbf{Target 1 (/200)}} & \multicolumn{4}{c|}{\textbf{Target 2 (/200)}} & \multicolumn{4}{c|}{\textbf{Target 3 (/200)}} \\ \cline{2-13} 
 & \multicolumn{4}{c|}{\textbf{(FA, MD)}} & \multicolumn{4}{c|}{\textbf{(FA, MD)}} & \multicolumn{4}{c|}{\textbf{(FA, MD)}} \\ \cline{2-13} 
 & \multicolumn{1}{l|}{\textbf{SPFL}} & \multicolumn{1}{l|}{\textbf{\textless{}32,5\textgreater{}}} & \multicolumn{1}{l|}{\textbf{\textless{}24,5,\textgreater{}}} & \multicolumn{1}{l|}{\textbf{\textless{}19,5\textgreater{}}} & \multicolumn{1}{l|}{\textbf{SPFL}} & \multicolumn{1}{l|}{\textbf{\textless{}32,5\textgreater{}}} & \multicolumn{1}{l|}{\textbf{\textless{}24,5,\textgreater{}}} & \multicolumn{1}{l|}{\textbf{\textless{}19,5\textgreater{}}} & \multicolumn{1}{l|}{\textbf{SPFL}} & \multicolumn{1}{l|}{\textbf{\textless{}32,5\textgreater{}}} & \multicolumn{1}{l|}{\textbf{\textless{}24,5,\textgreater{}}} & \multicolumn{1}{l|}{\textbf{\textless{}19,5\textgreater{}}} \\ \hline
\textbf{20} & \multicolumn{1}{c|}{(0,0)} & \multicolumn{1}{c|}{(0,0)} & \multicolumn{1}{c|}{(0,0)} & (0,0) & \multicolumn{1}{c|}{(8,0)} & \multicolumn{1}{c|}{(10,0)} & \multicolumn{1}{c|}{(12,0)} & (11,0) & \multicolumn{1}{c|}{(29,0)} & \multicolumn{1}{c|}{(30,0)} & \multicolumn{1}{c|}{(30,1)} & (30,101) \\ \hline
\textbf{10} & \multicolumn{1}{c|}{(0,0)} & \multicolumn{1}{c|}{(0,0)} & \multicolumn{1}{c|}{(0,0)} & (0,0) & \multicolumn{1}{c|}{(7,0)} & \multicolumn{1}{c|}{(8,1)} & \multicolumn{1}{c|}{(10,2)} & (12,2) & \multicolumn{1}{c|}{(30,0)} & \multicolumn{1}{c|}{(31,0)} & \multicolumn{1}{c|}{(31,2)} & (31,107) \\ \hline
\textbf{0} & \multicolumn{1}{c|}{(0,1)} & \multicolumn{1}{c|}{(0,1)} & \multicolumn{1}{c|}{(0,1)} & (0,6) & \multicolumn{1}{c|}{(7,5)} & \multicolumn{1}{c|}{(9,6)} & \multicolumn{1}{c|}{(8,6)} & (18,35) & \multicolumn{1}{c|}{(30,3)} & \multicolumn{1}{c|}{(29,4)} & \multicolumn{1}{c|}{(33,5)} & (33,99) \\ \hline
\textbf{-10} & \multicolumn{1}{c|}{(0,1)} & \multicolumn{1}{c|}{(0,0)} & \multicolumn{1}{c|}{(0,2)} & (0,10) & \multicolumn{1}{c|}{(3,5)} & \multicolumn{1}{c|}{(13,21)} & \multicolumn{1}{c|}{(12,22)} & (21,42) & \multicolumn{1}{c|}{(31,20)} & \multicolumn{1}{c|}{(31,26)} & \multicolumn{1}{c|}{(31,24)} & (31,106) \\ \hline
\end{tabular}%
}}
\end{table*}

\begin{table*}[!b]
\centering
\footnotesize
\caption{\footnotesize Comparison of resource utilization and power consumption for different WLs}
\label{tab:WL_Comp}
\renewcommand{\arraystretch}{1.3}
 \resizebox{\textwidth}{!}{
\begin{tabular}{|c|c|c|cc|cc|cc|cc|cc|}
\hline
\multirow{2}{*}{\textbf{No.}} & \multirow{2}{*}{\textbf{\begin{tabular}[c]{@{}c@{}}Range and Azimuth\\ Estimation WL\end{tabular}}} & \multirow{2}{*}{\textbf{\begin{tabular}[c]{@{}c@{}}Doppler Velocity\\ Estimation WL\end{tabular}}} & \multicolumn{2}{c|}{\textbf{BRAM}} & \multicolumn{2}{c|}{\textbf{DSP}} & \multicolumn{2}{c|}{\textbf{Flip Flops}} & \multicolumn{2}{c|}{\textbf{Look Up Tables (LUTs)}} & \multicolumn{2}{c|}{\textbf{Power}} \\ \cline{4-13} 
 &  &  & \multicolumn{1}{c|}{\textbf{Utilization}} & \textbf{Savings} & \multicolumn{1}{c|}{\textbf{Utilization}} & \textbf{Savings} & \multicolumn{1}{c|}{\textbf{Utilization}} & \textbf{Savings} & \multicolumn{1}{c|}{\textbf{Utilization}} & \textbf{Savings} & \multicolumn{1}{c|}{\textbf{Consumption}} & \textbf{Savings} \\ \hline
1 & SPFL & \multirow{4}{*}{SPFL} & \multicolumn{1}{c|}{1073} & NA & \multicolumn{1}{c|}{1655} & NA & \multicolumn{1}{c|}{144001} & NA & \multicolumn{1}{c|}{151712} & NA & \multicolumn{1}{c|}{6.59} & NA \\ \cline{1-2} \cline{4-13} 
2 & \textless{}32, 5\textgreater{} &  & \multicolumn{1}{c|}{818} & \textbf{24\%} & \multicolumn{1}{c|}{1163} & \textbf{30\%} & \multicolumn{1}{c|}{41001} & \textbf{71\%} & \multicolumn{1}{c|}{63382} & \textbf{58\%} & \multicolumn{1}{c|}{3.7} & \textbf{44\%} \\ \cline{1-2} \cline{4-13} 
3 & \textless{}24, 5\textgreater{} &  & \multicolumn{1}{c|}{638} & \textbf{40\%} & \multicolumn{1}{c|}{600} & \textbf{64\%} & \multicolumn{1}{c|}{35078} & \textbf{76\%} & \multicolumn{1}{c|}{44152} & \textbf{71\%} & \multicolumn{1}{c|}{3.3} & \textbf{50\%} \\ \cline{1-2} \cline{4-13} 
4 & \textless{}19, 5\textgreater{} &  & \multicolumn{1}{c|}{438} & \textbf{59\%} & \multicolumn{1}{c|}{374} & \textbf{77\%} & \multicolumn{1}{c|}{37924} & \textbf{74\%} & \multicolumn{1}{c|}{36481} & \textbf{76\%} & \multicolumn{1}{c|}{3.1} & \textbf{53\%} \\ \hline
\end{tabular}%
}
\end{table*}
\vspace{-0.2cm}
\subsection{Fixed-Point Architecture}
\begin{figure}[!t]
\vspace{-0.1cm}
    \centering
    \includegraphics[width=0.9 \linewidth]{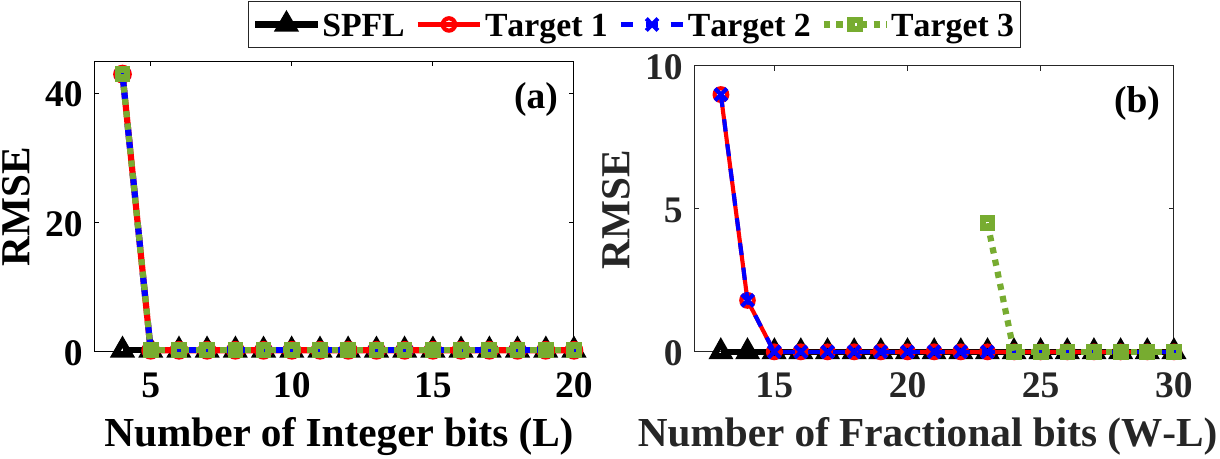}
    \vspace{-0.1cm}
    \caption{\footnotesize Average range and azimuth RMSE of 3 targets for different WL: (a) RMSE vs. number of integer bits, $L$, for 30 fractional bits, (b) RMSE vs. number of fractional bits for 5 integer bits.}
    \label{fig:rmse_WLSel}
\vspace{-0.6cm}
\end{figure}

We analyze the effect of WL on the functional correctness of the proposed hardware IPs. Ideally, the WL should be as small as possible, and fixed-point WL architectures are preferred over SPFL architectures due to lower resource utilization, execution time, and power consumption. We denote the fixed-point WL as \textless{}$W,L$\textgreater{} where $W$ is the total number of bits, and $L$ and $W-L$ are the number of bits used to represent the integer and fractional parts of the real number respectively. To identify the desired WL, we first use a sufficiently large number value for $W$ and vary the integer bits, $L$. This gives the minimum value $L$ needed to represent all numerical values accurately. As shown in Fig.~\ref{fig:rmse_WLSel} (a), a total of 5 integer bits are sufficient to ensure that the RMSE of range and azimuth estimation for all three targets is the same as that of the SPFL implementation. The results are averaged over SNRs ranging from -15 to 20 dB.
\begin{figure*}[!hb]
    \centering
    \includegraphics[scale = 0.45]{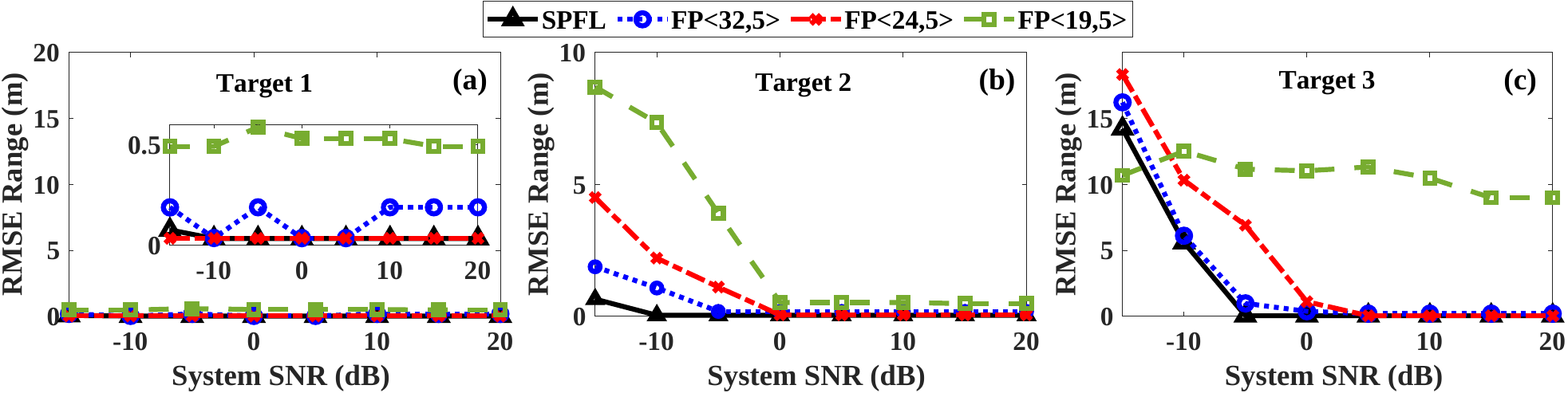}
    \vspace{-0.2cm}
     \caption{\footnotesize Functionality verification of fixed point architectures of range estimation for three targets.}
     \vspace{-0.2cm}
    \label{fig:rmse_FP_range}
\end{figure*}
\begin{figure*}[!hb]
\vspace{-0.2cm}
    \centering
    \includegraphics[scale = 0.45]{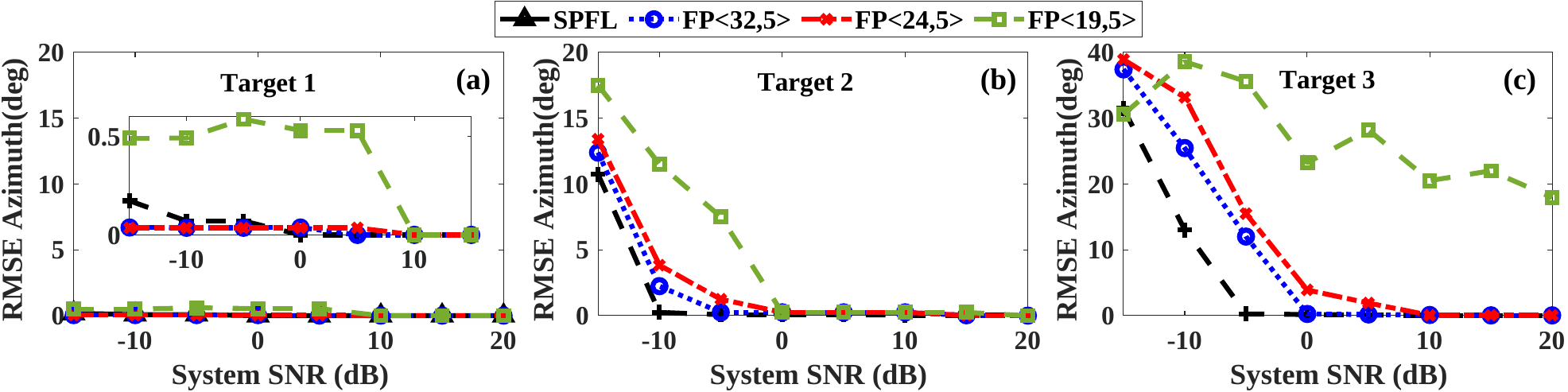}
     \vspace{-0.2cm}
     \caption{\footnotesize Functionality verification of fixed point architectures of azimuth estimation for three targets.}
     \vspace{-0.2cm}
    \label{fig:rmse_FP_azimuth}
\end{figure*}
\begin{figure*}[!hb]
    \centering
    \vspace{-0.2cm}
    \includegraphics[scale = 0.45]{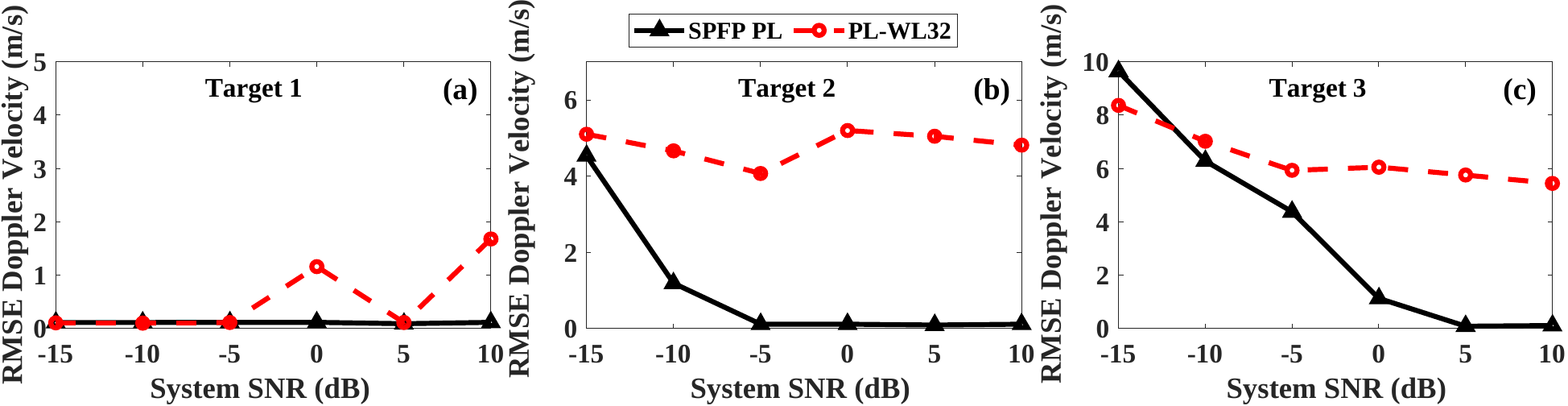}
     \vspace{-0.2cm}
     \caption{\footnotesize Functionality verification of fixed point architectures of Doppler velocity estimation for three targets.}
     \vspace{-0.2cm}
     \label{fig:rmse_FP_Doppler}
\end{figure*}
\begin{figure*}[!hb]
    \centering
    \vspace{-0.2cm}
    \includegraphics[scale = 0.45]{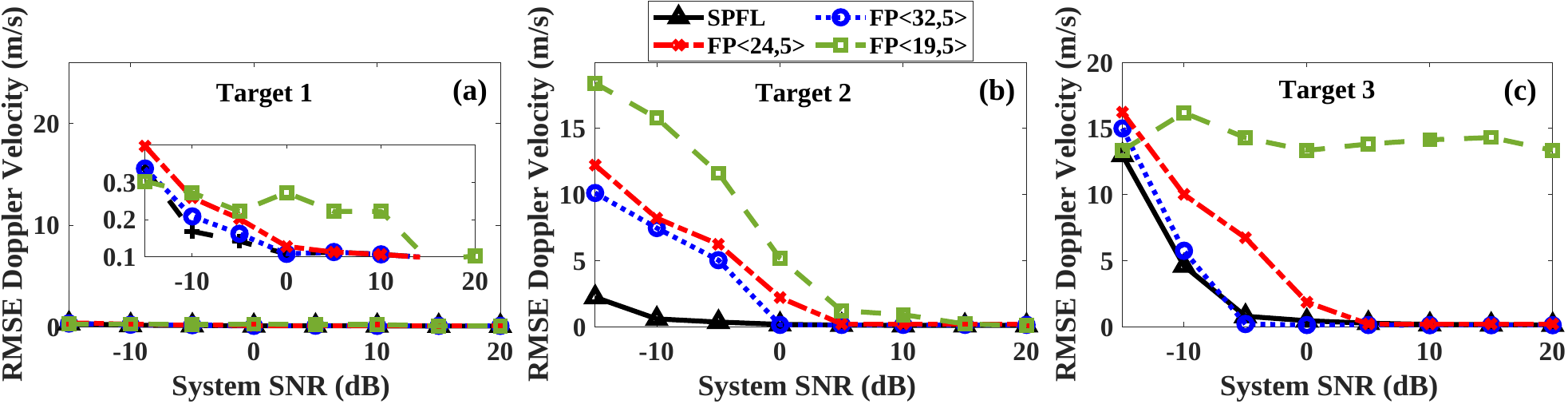}
     \vspace{-0.2cm}
     \caption{\footnotesize Functionality verification of SPFL architecture of Doppler velocity estimation for different fixed-point MF architectures.}
     \vspace{-0.2cm}
     \label{fig:rmse_FXP_Doppler}
\end{figure*}
Next, we fixed the number of integer bits to 5 and varied the $W$ to identify the minimum number of bits, $(W-L)$, needed to represent fractional numbers accurately. As shown in Fig.~\ref{fig:rmse_WLSel} (b), the value of $(W-L)$ is not the same for all the targets. The third target needs more bits than the first two targets. Since we use the same IP for detecting all three targets, we use the WL of \textless{}$24,5$\textgreater{} for MF IP. In the case of FFT and IFFT, we have used the IPs provided by AMD-Xilinx, which are optimized for SPFL WL. 

Next, we verify the functional correctness of the fixed-point hardware MF IPs for a wide range of SNR for the range and azimuth estimation of the first, second, and third targets. 
As shown in Fig.~\ref{fig:rmse_FP_range} and Fig.~\ref{fig:rmse_FP_azimuth}, the selected WL of \textless{}$24,5$\textgreater{} offers nearly the same RMSE as that of the SPFL architectures for all SNR under different target scenarios. In Table~\ref{tab:FA_MD_WL}, we compare the FA and MD metrics for different WLs and it can be observed that the selected WL of \textless{}$24,5$\textgreater{} offers nearly the same performance as that of the SPFL architectures for all SNRs under different target scenarios.

Further reduction in WL to \textless{}$19,5$\textgreater{} results in significant degradation in performance. The WL of  \textless{}$32,5$\textgreater{} has the same number of bits as that of SPFL, and both offer nearly identical RMSE. As discussed later, the former WL of \textless{}$32,5$\textgreater{} and \textless{}$24,5$\textgreater{} offer significant savings in resources, validating the importance of WL optimization on hardware. 
For MUSIC-based Doppler estimation, the matrix arithmetic involving EVD demands a large dynamic range; hence, the fixed-point architecture incurs large RMSE. Using the SPFL MF IP, we analyze the RMSE in the Doppler velocity estimation using MUSIC IPs of different WL. As shown in Fig.~\ref{fig:rmse_FP_Doppler}, the performance degrades significantly for the fixed-point architecture of the MUSIC and hence, SPFL WL is preferred for Doppler velocity estimation. Since MUSIC-based Doppler velocity estimation depends on the MF IP output, we compare the performance of SPFL MUSIC IP for different fixed-point architectures of MF IP. As shown in Fig.~\ref{fig:rmse_FXP_Doppler}, the WL of \textless{}$24,5$\textgreater{} does not affect the RMSE of Doppler velocity, thereby validating the selected WL of \textless{}$24,5$\textgreater{}. 

\section{Complexity Comparison}
\label{Sec:CC}
This section compares the resource utilization, power consumption, and execution time of various architectures obtained via WL and HSCD optimizations. In Table~\ref{tab:WL_Comp}, we compare the proposed architecture's FPGA resource utilization and power consumption for four different WLs. It can be observed that the SPFL architecture consumes the highest amount of block RAM (BRAM), embedded DSP multiplier units, flip-flops (FFs), and FPGA look-up tables (LUTs). The dynamic power consumption of the SPFL architecture is the highest. The proposed architecture with WL of \textless{}$32,5$\textgreater{} offers identical functional performance as that of SPFL architecture but significantly reduces resource utilization and power consumption. Further reduction of around 40-76\% in resource utilization is achieved using the WL of \textless{}$24,5$\textgreater{} without compromising functional accuracy for all three target detection, as shown in the table. In the case of two target detection, we can further reduce the WL up to \textless{}$19,5$\textgreater{} to gain 59-76\% savings in resources and 53\% in power consumption with a slight degradation in functional accuracy.

The HSCD divides the various tasks between PL/hardware (i.e., FPGA) and PS/software (i.e., ARM Processor). When algorithms are mapped on the FPGA, there are the benefits of reduced execution time for data-intensive computations and reconfigurability at the cost of increased power consumption and computational resources. We compare the time required to estimate the range and azimuth of the single target using one packet and the Doppler velocity of the single target using 100 packets. We assume all the packets are available in memory, so there is no time overhead for data communications between the PS and PL. As shown in Table~\ref{tab:EXT_Comp}, the FPGA-based implementation results in a huge improvement in execution time with accelerator factors of 204 and 14.5 for range-azimuth and Doppler velocity estimation, respectively.
\begin{table}[!t]
\centering
\caption{\footnotesize Comparison of execution time for RSP with SPFL WL}
\label{tab:EXT_Comp}
\renewcommand{\arraystretch}{1.2}
\resizebox{\columnwidth}{!}{%
\begin{tabular}{|c|c|cc|c|}
\hline
\multirow{2}{*}{\textbf{Target Parameter}} & \multirow{2}{*}{\textbf{Packets}} & \multicolumn{2}{c|}{\textbf{Execution Time (Seconds)}} & \multirow{2}{*}{\textbf{\begin{tabular}[c]{@{}c@{}}Acceleration\\ Factor\end{tabular}}} \\ \cline{3-4}
 &  & \multicolumn{1}{c|}{\textbf{PS}} & \textbf{PS+PL} &  \\ \hline
Range and Azimuth & 1 & \multicolumn{1}{c|}{5.3} & 0.026 & \textbf{204} \\ \hline
Doppler Velocity & 100 & \multicolumn{1}{c|}{1.1} & 0.077 & \textbf{14.5} \\ \hline
\end{tabular}%
}
\vspace{-0.4cm}
\end{table}
\begin{table*}[!t]
\centering
\caption{\footnotesize Comparison of acceleration factor, resource utilization, and power consumption for different HSCD configurations}
\label{tab:HACD_Comp}
\renewcommand{\arraystretch}{0.98}
\resizebox{0.8\linewidth}{!}{
\footnotesize
\begin{tabular}{|c|c|c|c|cccc|c|}
\hline
\textbf{No.} &
  \textbf{Blocks in PL} &
  \textbf{WL} &
  \textbf{\begin{tabular}[c]{@{}c@{}}Acceleration \\ Factor (AF)\end{tabular}} &
  \multicolumn{1}{c|}{\textbf{BRAM}} &
  \multicolumn{1}{c|}{\textbf{DSP}} &
  \multicolumn{1}{c|}{\textbf{Flip Flops}} &
  \textbf{LUTs} &
  \textbf{\begin{tabular}[c]{@{}c@{}}Dynamic\\ Power (W)\end{tabular}} \\ \hline
1 &
  NA &
  SPFL &
  \textbf{1} &
  \multicolumn{4}{c|}{NA} &
  NA \\ \hline
2 &
  Doppler Velocity &
  SPFL &
  \textbf{1.01} &
  \multicolumn{1}{c|}{262} &
  \multicolumn{1}{c|}{350} &
  \multicolumn{1}{c|}{28923} &
  40800 &
  3.2 \\ \hline
3 &
  Range and Azimuth &
  SPFL &
  \textbf{76} &
  \multicolumn{1}{c|}{685} &
  \multicolumn{1}{c|}{732} &
  \multicolumn{1}{c|}{80849} &
  60067 &
  3.9 \\ \hline
4 &
  \begin{tabular}[c]{@{}c@{}}Doppler Velocity,\\ Range and Azimuth\end{tabular} &
  SPFL &
  \textbf{104} &
  \multicolumn{1}{c|}{1073} &
  \multicolumn{1}{c|}{1655} &
  \multicolumn{1}{c|}{144001} &
  151712 &
  6.6 \\ \hline
5 &
  Range and Azimuth &
  \textless{}24,5\textgreater{} &
  \textbf{94.6} &
  \multicolumn{1}{c|}{381} &
  \multicolumn{1}{c|}{300} &
  \multicolumn{1}{c|}{24703} &
  21336 &
  2.8 \\ \hline
6 &
  \begin{tabular}[c]{@{}c@{}}Doppler Velocity,\\ Range and Azimuth\end{tabular} &
  \textless{}24,5\textgreater{} &
  \textbf{120} &
  \multicolumn{1}{c|}{638} &
  \multicolumn{1}{c|}{600} &
  \multicolumn{1}{c|}{35078} &
  44152 &
  3.3 \\ \hline
7 &
  \begin{tabular}[c]{@{}c@{}}Doppler Velocity,\\ Range and Azimuth (2)\end{tabular} &
  \textless{}24,5\textgreater{} &
  \textbf{149} &
  \multicolumn{1}{c|}{800} &
  \multicolumn{1}{c|}{1152} &
  \multicolumn{1}{c|}{60956} &
  98122 &
  5.3 \\ \hline
\end{tabular}}\\
\noindent
\vspace{2mm}
\footnotesize{*Row 7 corresponds to an alternative serial-parallel configuration with 2 MF IPs.}
\vspace{-0.5cm}
\end{table*}
\begin{figure*}[!b]
    \centering
    \includegraphics[scale = 0.46]{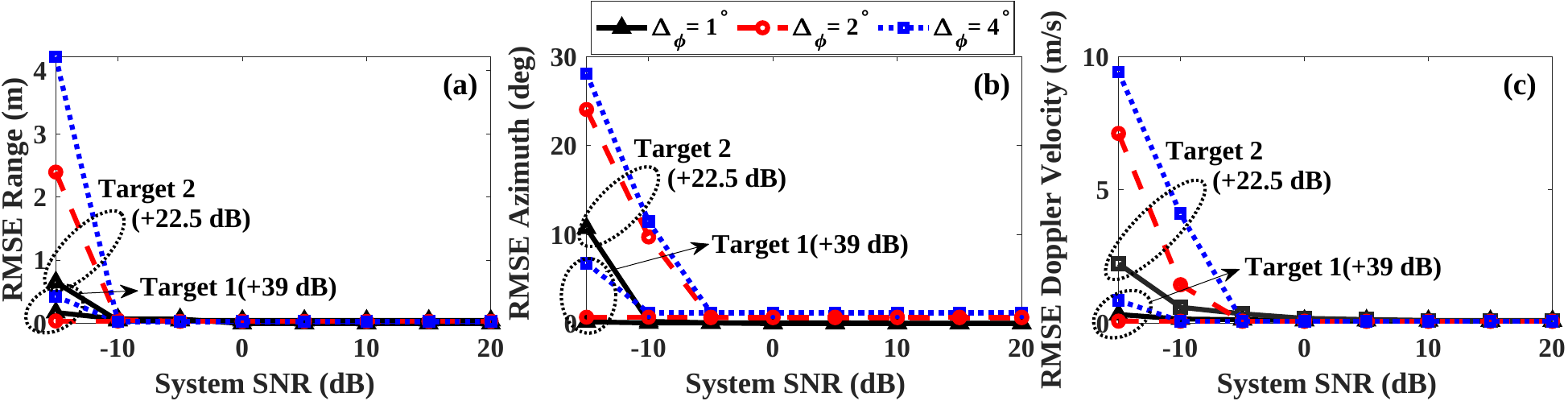}
    \vspace{-0.2cm}
    \caption{\footnotesize RMSE for different angular precision values for varying system SNR}
    \label{fig:rmse-ap}
\end{figure*}
\begin{figure*}[!b]
    \centering
    \vspace{-0.3cm}
    \includegraphics[scale = 0.45]{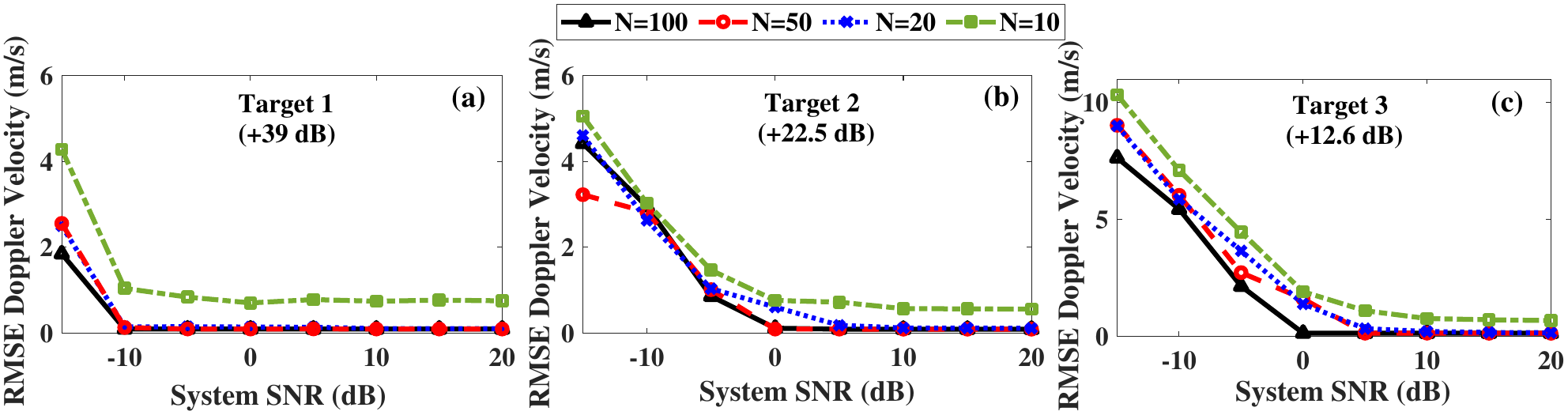}
    \vspace{-0.2cm}
    \caption{\footnotesize RMSE with MUSIC for different numbers of packets for varying system SNR}
    \label{fig:rmse-packet}
\end{figure*}

Next, we consider the HSCD of the end-to-end RSP, which includes the transmitter, wireless channel, and target modeling followed by sequential reception and processing of 100 packets at the receiver using the proposed RSP architecture. As shown in Table~\ref{tab:HACD_Comp}, we compare the performance of various configurations with the configuration in Row 1, where complete processing is done in PS. Such a configuration does not require the FPGA; hence, the resource utilization is mentioned as not applicable (NA). However, this configuration incurs the highest execution time due to the sequential nature of the software processing; hence, the acceleration factor (AF) is mentioned as 1. Note that this is despite using an ARM Cortex A53 quad-core processor operating at 1.2 GHz compared to 300 MHz in FPGA. In addition, we have deployed the PYNQ operating system on PS, which allows parallelism using multiple cores.

In all the configurations discussed next, CLEAN and peak search algorithms are realized in PS. In the second configuration (Row 2), we have moved the MUSIC-based Doppler velocity estimation to PL, which only slightly improved the execution time with an AF of 1.01. However, there is a significant increase in FPGA resource utilization and power consumption. In the third configuration (Row 3), we have moved the MF and FFT-based range and azimuth estimation in PL. The rest of the processing, including Doppler velocity estimation, is realized in PS. This configuration offers a significant AF of 76 over the first configuration. Similarly, in the fourth configuration (Row 4), all estimation algorithms are realized in PL, resulting in an AF of 104. Though the third and fourth configurations offer good acceleration, they incur high resource utilization and power consumption. This can be reduced significantly by using fixed-point architectures, as shown in Rows 5 and 6 of Table~\ref{tab:HACD_Comp}. Further improvement in AF and reduction in resource utilization is possible by exploring fixed-point realization of the MUSIC. However, it is challenging due to the large number of packets that demand auto-covariance and EVD of high dimension signal. This demands more efforts towards the fixed-point realization of MUSIC for 3D RSP, especially for large $N$.

In the existing architecture, the precision of azimuth estimation ($\Delta \phi$) is $1^{\circ}$, meaning there are 181 candidate angles when the search space is from $-90^{\circ}$ to $+90^{\circ}$. For each angle, we need to perform the MF operation in Steps 2 and 3 in Section~\ref{Sec:RSPF}. Due to resource constraints, we can not have a fully parallel architecture with 181 MF IPs. In this direction, we explore a new configuration of the proposed architecture by having two MF IPs resulting in the serial-parallel configuration as shown in Row 7 of Table~\ref{tab:HACD_Comp}. As expected, a significant gain in AF is achieved with a slight increase in resource utilization and power consumption. Similar architecture can be explored for FFT operations which are currently being done sequentially by using a single IP shared across all antennas. When we compare architectures in Rows 4 and 7, the architecture in Row 7 offers a 1.5 times higher AF, needs fewer resources, and consumes less power. This validates the importance of careful HSCD, WL optimization, and serial-parallel architectures. 

\vspace{-0.4cm}
\section{Reconfigurable Architecture}
\vspace{-0.2cm}
\label{Sec:ReconfA}


In the previous sections, we assumed $\Delta \phi$ to be $1^{\circ}$ and $|I|=181$ (for $\phi_i$ spanning $-90^{\circ}$ to $+90^{\circ}$). In many application scenarios, we may however need lower precision in the beginning and finer precision later. For instance, base stations may need to identify clusters of users quickly using a coarse sectoral search before estimating the accurate azimuth of each user in a given cluster. The proposed architecture enables on-the-fly configuration of $\Delta \phi$, which allows significant improvement in execution time. As shown in Fig.~\ref{fig:rmse-ap}, the increase in $\Delta \phi$ does not affect the RMSE of range and Doppler velocity however, there is a marginal degradation of the RMSE for azimuth.  On the other hand, the execution time for $\Delta \phi = 2^{\circ}$ and $\Delta \phi = 4^{\circ}$ is 1.96 and 3.75 times lower than that of $\Delta \phi = 1^{\circ}$. Also, the dynamic power consumption for $\Delta \phi = 2^{\circ}$ and $\Delta \phi = 4^{\circ}$ is 15\% and 20\% times lower than that of $\Delta \phi = 1^{\circ}$.

\begin{figure}[!h]
    \centering
    \includegraphics[scale = 0.4]{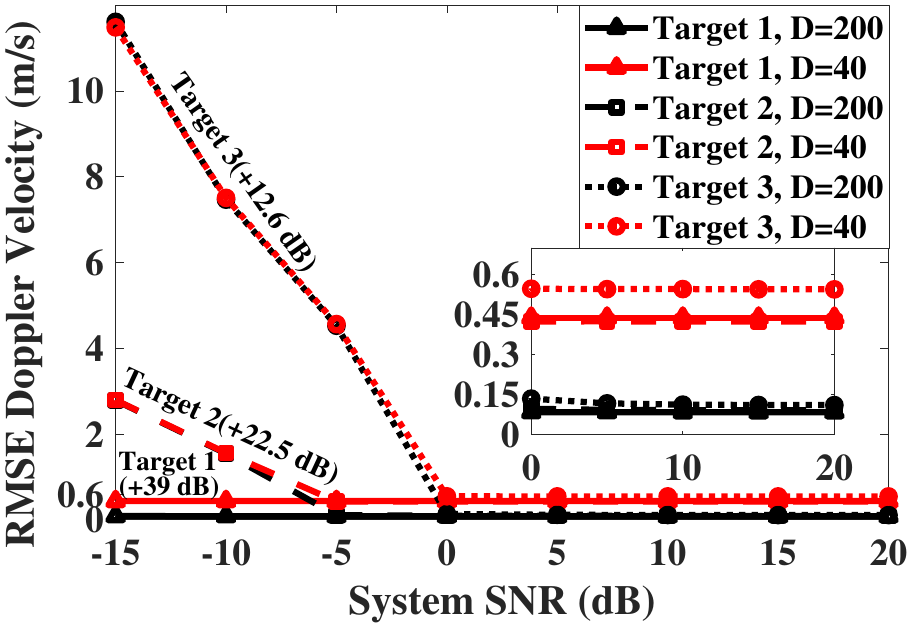}
    \vspace{-0.2cm}
    \caption{\footnotesize The effect of the number of Doppler bins, $D$, on the RMSE of the Doppler velocity for three targets with Doppler velocity spans from $-30m/s$ to $30m/s$.}
    \label{fig:reconfDD}
\end{figure}

Next, we consider the on-the-fly configuration of a second parameter - the number of packets. Based on clutter conditions (for example, clutter differs significantly in highway and urban scenarios), the number of packets in a coherent interval may need to be dynamically adjusted. A fewer number of packets (shorter coherent intervals) can significantly reduce the estimation time. We consider the effect of the number of packets on the RMSE of Doppler velocity estimation via MUSIC for different SNRs.
As shown in Fig.~\ref{fig:rmse-packet}, the Doppler RMSE increases with a fall in SNR. This is most significant for the third target case, which has the lowest signal strength (Fig.~\ref{fig:rmse-packet}c). Furthermore, the reduction in RMSE is insignificant after a certain number of packets. On the other hand, execution time is reduced by 9, 5, and 2 factors for 10, 20, and 50 packets when compared to 100 packets. Thus, we can reduce the execution time significantly by reducing the number of packets with a slight degradation in the RMSE. Many applications do not need accurate Doppler velocity but need to know whether the target velocity is within a specific range. For such applications, we can tune the number of Doppler bins, $D$, in the MSG block of MUSIC, which controls the resolution of Doppler velocity estimation. As shown in Fig.~\ref{fig:reconfDD}, the maximum error is $0.15m/s$ and $0.6m/s$ when $D$ is 200 and 40, respectively. On the other hand, the execution time is reduced by a factor of 2.5 for $D=40$. Since MUSIC IP is active once per target compared to MF IP, which is activated once per packet, there are no significant savings in the total power of RSP due to the reduced number of Doppler bins. 

\section{FFT versus MUSIC}
\label{Sec:FFTvsMUSIC}
\begin{figure}[!t]
    \centering
     \vspace{-0.3cm}
    \includegraphics[scale = 0.35]{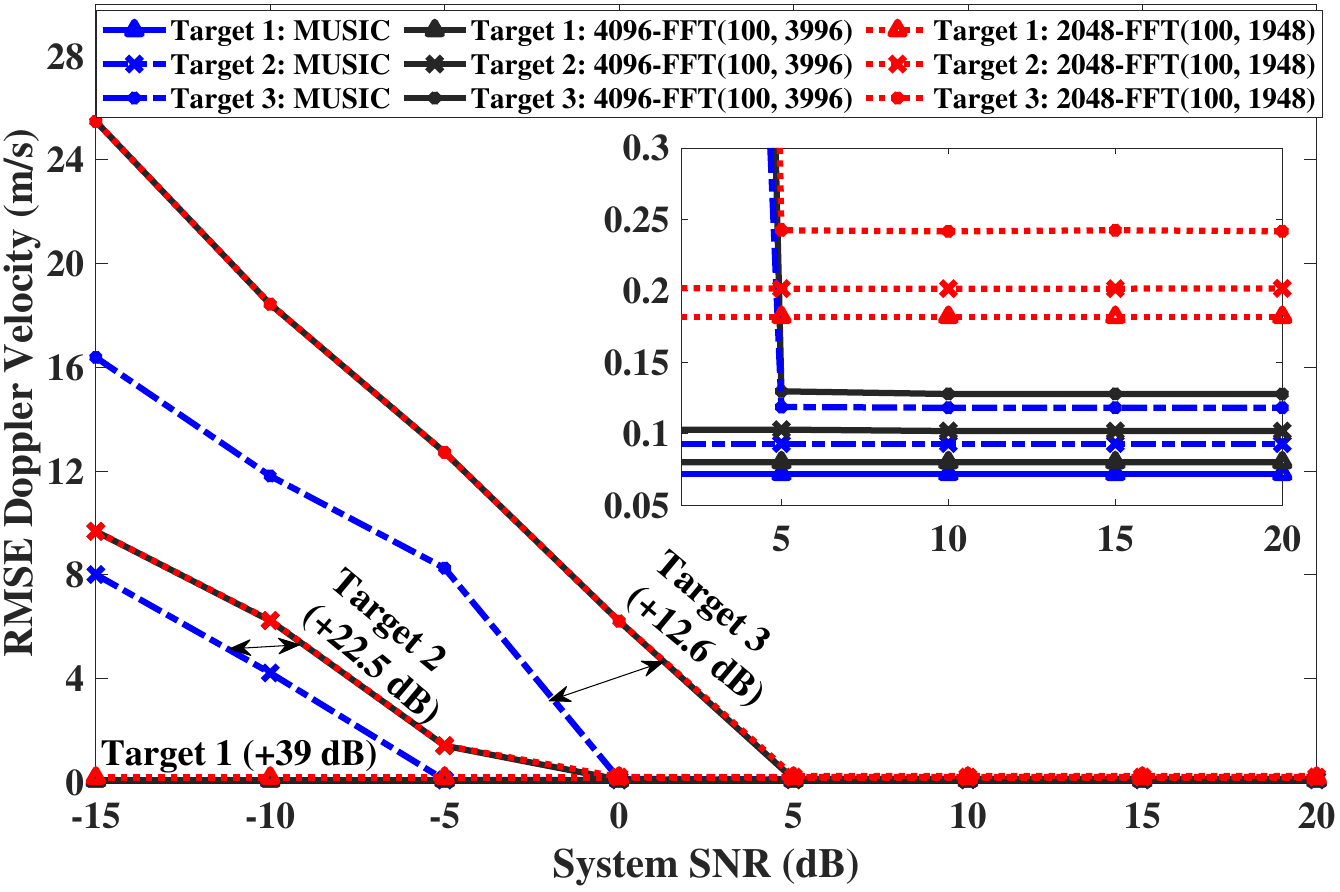}
    \vspace{-0.2cm}
    \caption{\footnotesize Doppler velocity RMSE comparison between MUSIC and FFT after zero padding for three targets with 100 packets.}
     \vspace{-0.1cm}
    \label{fig:music_fftzp}
\end{figure}

We compare the Doppler RMSE between MUSIC and FFT-based approaches using an identical number of packets (100). Further, the Doppler velocity bin size/precision for both methods is set to be identical by adjusting the bin size in the pseudo-spectrum of MUSIC and through zero padding/sinc interpolation of the FFT spectrum to 2048/4096 samples. In Fig.\ref{fig:music_fftzp}, the notation 4096-FFT (number of packets:100, zero-padding:3996) refers to 4096-FFT processing for 100 packets after zero-padding 3996 time samples. We observe from the results that the performance of MUSIC and 4096-FFT (100, 3996) is nearly identical for high SNRs (5dB and above) for all three targets whereas 2048-FFT(number of packets:100, zero-padding:1948) is slightly degraded in accuracy. However, at lower SNRs, MUSIC outperforms FFT for the weaker targets (2 and 3).

\begin{figure}[!t]
 \vspace{-0.3cm}
    \centering
    \includegraphics[scale = 0.38]{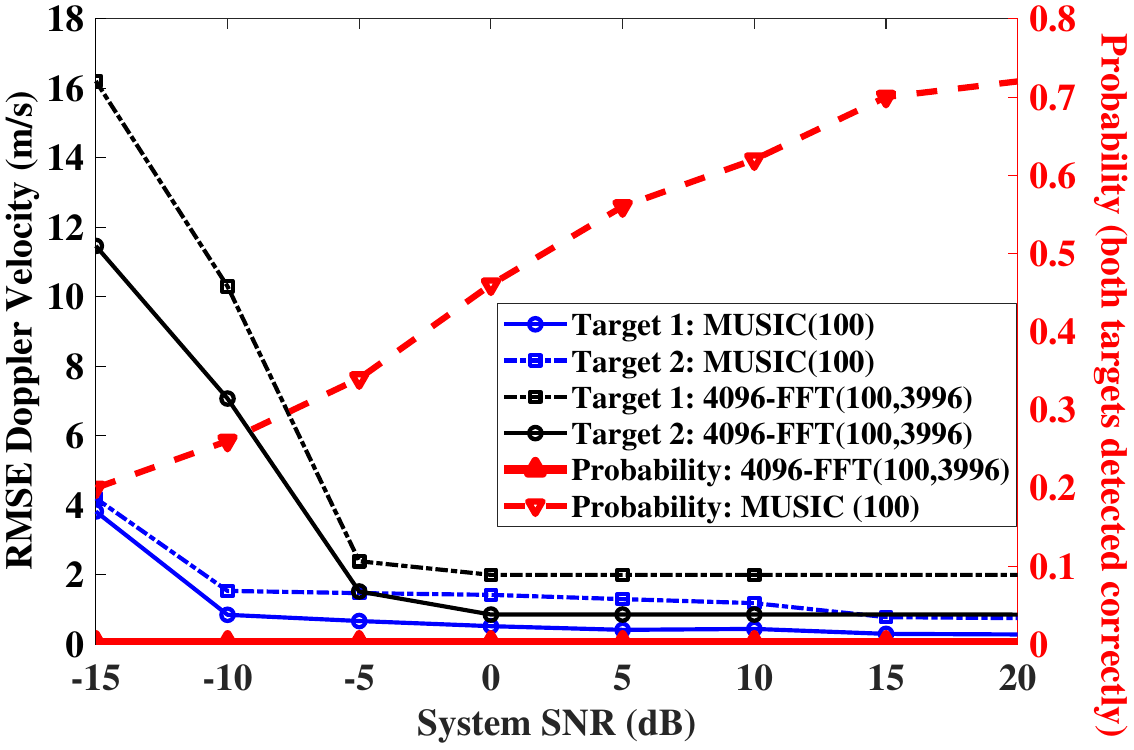}
    \vspace{-0.2cm}
    \caption{\footnotesize Doppler velocity RMSE with 100 packets for a target and a static clutter within the same range-azimuth bin along with the probability of both velocities being detected correctly (special case).}
     \vspace{-0.3cm}
    \label{fig:super_reso}
\end{figure}

Next, we consider a special case with two targets of closely spaced Doppler velocities within the same range-azimuth cell. Here, one of the targets is modeled as a static clutter with zero Doppler velocity. The difference in the Doppler velocity of the clutter scatterer and the target is less than or equal to 4 m/s. The RMSE for both targets and the probability that each target is detected correctly is shown for MUSIC and 4096-FFT (100, 3996) in Fig.\ref{fig:super_reso}. We observe that at 20 dB SNR, MUSIC can detect two Doppler velocities with a probability of 0.7. The probability degrades to 0.2 with decreasing SNR. However, since FFT produces only a single peak, the probability of detecting both scatterers is zero. The results in Fig.\ref{fig:super_reso} clearly show that despite the identical coherent processing intervals and precision of both algorithms, MUSIC achieves lower Doppler velocity error due to super-resolution capabilities.

\begin{table}[!b]
\footnotesize
\centering
\caption{\footnotesize Hardware complexity and latency comparison for MUSIC and 4096-FFT(100,3996) with 100 Packets}
\label{tab:musicVSfft}
\resizebox{1.03\columnwidth}{!}{
\begin{tabular}{|c|c|c|c|c|c|cc|}
\hline
\multirow{2}{*}{\textbf{Algorithm}} &
  \multirow{2}{*}{\textbf{BRAM}} &
  \multirow{2}{*}{\textbf{DSP}} &
  \multirow{2}{*}{\textbf{FF}} &
  \multirow{2}{*}{\textbf{LUT}} &
  \multirow{2}{*}{\textbf{\begin{tabular}[c]{@{}c@{}}Dynamic \\ Power\\ (W)\end{tabular}}} &
  \multicolumn{2}{c|}{\textbf{Acceleration Factor (AF)}} \\ \cline{7-8} 
 &
   &
   &
   &
   &
   &
  \multicolumn{1}{c|}{\textbf{\begin{tabular}[c]{@{}c@{}}Doppler \\ Processing\end{tabular}}} &
  \textbf{\begin{tabular}[c]{@{}c@{}}Complete \\ RSP\end{tabular}} \\ \hline
\textbf{MUSIC} &
  262 &
  350 &
  \multicolumn{1}{l|}{28923} &
  40800 &
  3.23 &
  \multicolumn{1}{c|}{1} &
  1 \\ \cline{1-2} \cline{3-8} 
\textbf{4096-FFT(100,3996)} &
  21 &
  63 &
  19875 &
  17576 &
  2.523 &
  \multicolumn{1}{c|}{31.53} &
  1.026 \\ \hline
\end{tabular}
}
\end{table}

As shown in Table.\ref{tab:musicVSfft}, the FFT algorithm requires much lower FPGA resources - in terms of BRAM, DSP, FF, and LUT - and computational time than MUSIC. The higher BRAM and DSP utilization in MUSIC can be attributed to the eigenvector decomposition operation on a large-size matrix. FFT, on the other hand, comprises linear operations and its efficient implementation exploiting the butterfly architecture results in a latency speed of 30 times faster than MUSIC. However, the overall AF for the whole RSP is still only 1.026 times greater for FFT than MUSIC. This is because the bulk of the processing time is still used for range-azimuth processing. 

\section{ISAC Performance Analysis}
{In this section, we evaluate the communication performance of our integrated ISAC system, discussed in Section-\ref{Sec:RSPArch}. G, and compare it with the standard IEEE 802.11ad in terms of BER and throughput.}
 \label{Sec:comm_performance}
\subsection{Simulation Model}
\begin{figure}[!h]
    \centering
    \includegraphics[scale = 0.28]{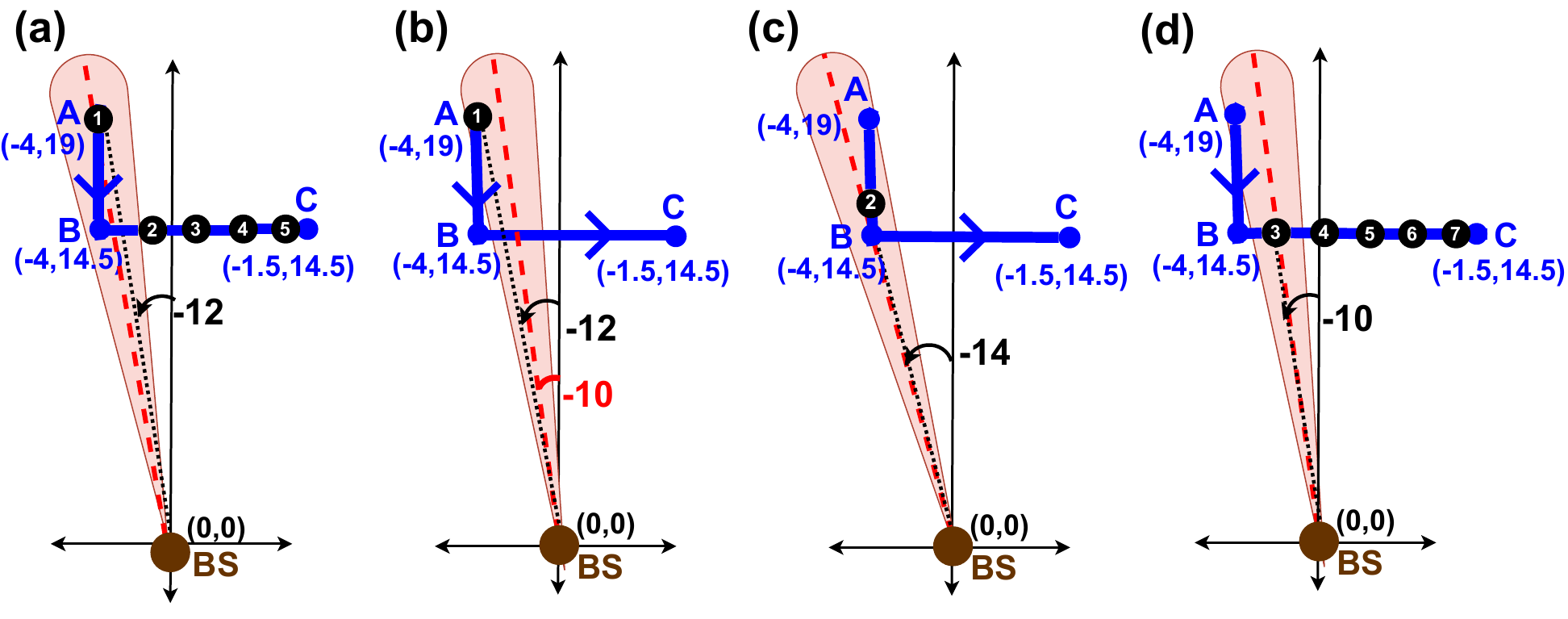}
    \vspace{-0.3cm}
    \caption{\footnotesize Beam alignment stages between the MU and ISAC BS with RSP angular precision of (a) 1$^\circ$ and (b-d) 4$^\circ$. The MU follows an L-shaped trajectory A-B-C for 1 second. (The figure is not drawn to scale.) }
    \label{fig:beam_s1}
\end{figure}

{We consider a point-size mobile user with an RCS of 10$ m^2$ moving along an L-shaped trajectory with an average velocity of 10m/s for 1 second. As shown in Fig.\ref{fig:beam_s1}, the BS is placed at the origin (0,0), and the MU travels along the path A-B-C indicated in blue. At time $t=0$, stage-1 is commenced, which marks the initiation of the beam alignment between BS and MU. For an ISAC scenario, the ISAC BS determines the AoA of the MU through RSP, whereas the MU finds the best beam towards BS using beam search. The BS is equipped with 32 antennas and localizes the MU via digital beamforming, offering tunable angular precisions depending upon the desired estimation accuracy. Since the MU is modeled with only 4 antennas and fewer beams, the stage-1 duration is largely determined by RSP time.}

{ Figure\ref{fig:beam_s1} illustrates the number of beam alignments required between the BS and MU at various stages in the trajectory. The actual angle of MU to the BS is highlighted in black, whereas the angle to which the beam is steered by BS towards MU is highlighted in red. Figure \ref{fig:beam_s1}(a) indicates the number of stage-1 for an RSP angular precision of 1$^\circ$. At point A, the MU is at an angle of -12$^\circ$ from the BS. Due to fine angular search, the AoA of the MU is estimated accurately, and the beam towards MU is steered very close to this initial AoA. As a result, the MU remains within the BS's beam for a long duration, leading to few requirements of beam alignments. Figure \ref{fig:beam_s1}(b-d) illustrates the same for 4$^\circ$ angular precision. As shown in Fig.\ref{fig:beam_s1}(b), initially, at point A, the beam is aligned to -10$^\circ$,  which is farther from the original angle due to the coarse search. Thus, the MU soon moves out of the beam, leading to the requirement of additional beam alignment stages around the turn at point B as shown in Fig.\ref{fig:beam_s1}(c) and Fig.\ref{fig:beam_s1}(d). This results in more frequent misalignments between BS and MU.}

\vspace{-0.2cm}
\subsection{Communication Link Metrics}
\begin{figure*}[!h]
    \centering
    \includegraphics[scale = 0.36]{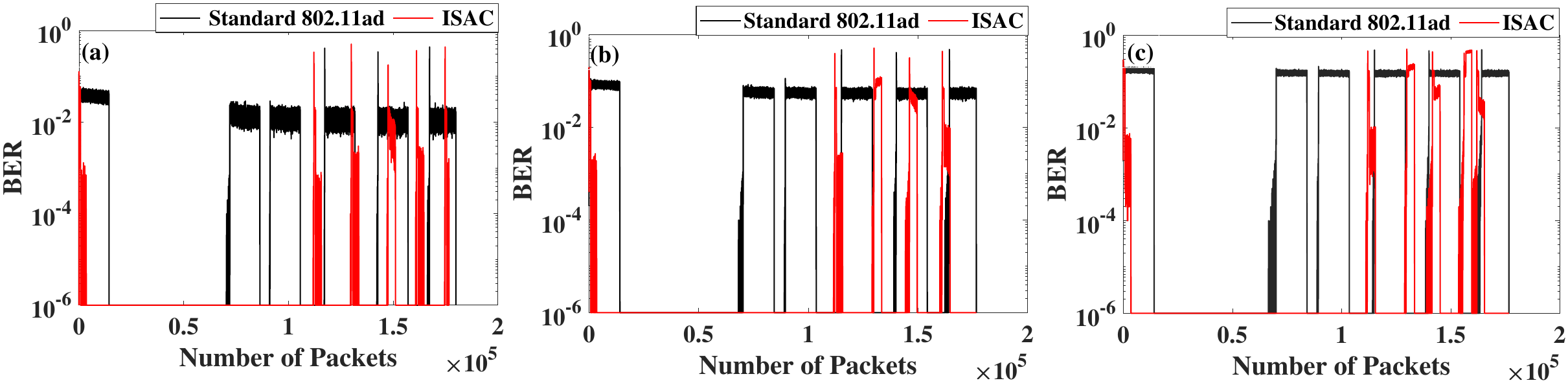}
    \vspace{-0.3cm}
    \caption{\footnotesize BER of standard IEEE 802.11ad and ISAC under (a) Friis' free space and Rician channel conditions for (b) J=7dB and (c) J=2dB for 1 second. }
    \label{fig:ber_channel}
\end{figure*}

{We analyze the performance of the above simulation model in terms of BER and throughput. Fig.\ref{fig:ber_channel} compares the BER of standard 802.11ad protocol with ISAC for the L-shaped trajectory for 1 second for different channels. The BER is plotted against the number of transmitted packets, which is analogous to time. Here, the BER is high during stage-1 (beam alignment stage between BS and MU) and drops during stage-2 (BS and MU are perfectly aligned). The ISAC system outperforms the standard 802.11ad due to its shorter stage-1 duration, facilitated by high-speed RSP at BS, compared to the lengthy beam training procedure required for standard 802.11ad processing. Table.\ref{Tab:throughput_std_isac} evaluates the throughput for different BS architectures and channel conditions over the 1-second duration of the above trajectory. We use an adaptive modulation scheme for high SNR conditions for enhanced throughput. Based on the downlink signal strength, the BS can switch between QPSK modulation for lower SNR and 16-QAM modulation for high SNR. As shown, ISAC provides higher throughput as compared to standard 802.11ad across all channel conditions.}

{We assess the performance for Friis' free space channel and Rician channel conditions with Rician factors of 7dB (weak NLOS component) and 2dB (strong NLOS component). Figure.\ref{fig:ber_channel}(a) shows a BER of up to 0.05 during stage 1 under Friis' free space channel, which degrades to 0.1 in Fig.\ref{fig:ber_channel}(b) and 0.2 in Fig.\ref{fig:ber_channel}(c) in Rician channel conditions due to the presence of multipath. Similarly, as shown in Table.\ref{Tab:throughput_std_isac}, the throughput drops in the case of Rician channel (J=2dB) as compared to the Friis' free space channel.}

\begin{figure*}[!h]
    \centering
    \includegraphics[scale = 0.34]{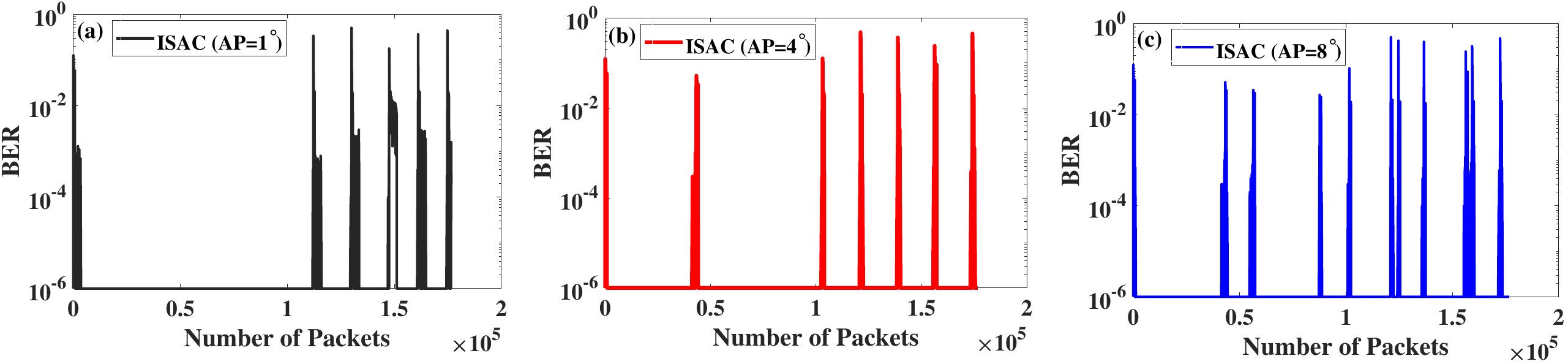}
    \vspace{-0.3cm}
    \caption{\footnotesize BER comparison for ISAC with different RSP azimuth precisions for L-shaped trajectory of MU for 1 second. }
    \label{fig:ber_AP}
\end{figure*}

{Figure \ref{fig:ber_AP} compares the BER rate across different ISAC angular search precisions. Even though 1$^\circ$ angular precision requires less frequent beam re-alignment stages as compared to 4$^\circ$, both configurations offer nearly identical throughput of 3.1 Gbps, as shown in Table.\ref{Tab:throughput_std_isac}. This is due to the faster RSP time with 4$^\circ$ precision as compared to 1$^\circ$ precision, leading to a shorter stage-1 duration. However, the 8$^\circ$ precision requires an even larger number of beam re-alignments, as shown in Fig.\ref{fig:ber_AP}(c), which causes the throughput to drop significantly, to 2.64 Gbps, as shown in Table.\ref{Tab:throughput_std_isac}.}


\begin{table}[!h]
\caption{\footnotesize Throughput comparison of standard IEEE 802.11ad and ISAC under different channel conditions for L-shaped trajectory with adaptive modulation scheme.}
\label{tab:jrc_throughput}
\resizebox{\linewidth}{!}{
\begin{tabular}{|c|c|ccc|}
\hline
\multirow{2}{*}{\textbf{Architecture}} &
  \multirow{2}{*}{\textbf{\begin{tabular}[c]{@{}c@{}}Angle precision \\ (degrees)\end{tabular}}} &
  \multicolumn{3}{c|}{\textbf{Throughput (Gbps)}} \\ \cline{3-5} 
                               &      & \multicolumn{1}{c|}{\textbf{Friis’ free space}} & \multicolumn{1}{c|}{\textbf{Rician (J=7dB)}} & \textbf{Rician (J=2dB)} \\ \hline

\textbf{Standard 802.11ad}     & -    & \multicolumn{1}{c|}{2.442}                      & \multicolumn{1}{c|}{2.414}                   & 2.3                   \\ \hline

\multirow{3}{*}{\textbf{ISAC}} & 1 & \multicolumn{1}{c|}{3.11}              & \multicolumn{1}{c|}{3.103} &2.99 \\ \cline{2-5}

                               & 4 & \multicolumn{1}{c|}{3.16}                      & \multicolumn{1}{c|}{3.101}                       &  3.05                       \\ \cline{2-5}
                               & 8 & \multicolumn{1}{c|}{2.635}                     & \multicolumn{1}{c|}{2.54}                        &    2.395                     \\ \hline
\end{tabular}}
\label{Tab:throughput_std_isac}
\end{table}

\section{Conclusions and Future Directions}
\label{sec:Conclusion}
In this work, we present a reconfigurable digital baseband RSP accelerator capable of three-dimensional target detection along the range, azimuth, and Doppler velocity. 
We have proposed novel IPs for hardware implementations of range-azimuth processing through matched filtering and digital beamforming; multiple target detection using CLEAN; and Doppler estimation using MUSIC algorithms on Zynq MPSoC. We have innovated a low-complexity framework for two-dimensional range-azimuth processing by exploiting the inherent redundancy in the data processing operations. We have benchmarked the functional correctness of our novel hardware IPs with the implementation of the algorithms in DPFL WL and demonstrated very good radar detection metrics in terms of low false alarms and missed detections along with low RMSE (below 1\%) for the estimation of range, Doppler and azimuth for multiple target scenarios. Proposed fixed-length architecture offers resource and dynamic power savings of the order of 40-76\%, and 50\%, respectively. We examine HSCD between PL and PS on the MPSoC and show that resource utilization on PL trades off with execution time; with an optimized architecture realized with Fourier-based range-azimuth processing on the PL and MUSIC-based Doppler estimation on PS. This fixed-point architecture results in an overall acceleration of 120 times with respect to the complete RSP on the PS. Further, the framework is made reconfigurable based on the selection of key parameters, such as the azimuth and Doppler precision, and the number of packets in the coherent processing interval, which makes it suitable for operation under different real-world scenarios with varied clutter and target conditions.{ We also integrate the reconfigurable RSP accelerator with communication PHY to demonstrate an end-to-end ISAC system on Zynq MPSoC, which offers superior communication link metrics over the standard 802.11ad.}  \balance

Some important points to note: First, RSP hardware IPs have been realized for a co-designed 802.11ad-based ISAC with Golay-based transmit waveforms. However, these IPs can be modified with minimum changes in the architecture for other types of ISAC waveforms such as the orthogonal time-frequency space (OTFS) characterized by a unique delay-Doppler signal structure \cite{gaudio2020effectiveness, wu2021otfs, gong2023super}. Only the pre-computed 1D-FFT sequences in \textbf{Step.1.} of the RSP would have to be modified for this purpose. Further, the reconfigurable nature of the RSP accelerator could be potentially exploited for on-the-fly changes in the radar transmit waveform. Second, the current RSP assumes a fixed threshold for the detection of the targets; while real-world conditions require an adaptive threshold. In future work, we will implement adaptive threshold selection based on constant false alarm rate (CFAR).
\section*{Appendix-I: 2D CLEAN versus 3D CLEAN}
We implement the CLEAN algorithm in the processor on MPSoC in two modes and compare the execution time for each: (1) 2-D: CLEAN performed after \emph{radar data square} is processed along range and azimuth to unmask the weaker targets using two-dimensional point spread response (PSR), followed by Doppler processing using MUSIC/fast Fourier transform (FFT); (2) 3-D: CLEAN performed after \emph{radar data cube} is processed along the range, azimuth, and Doppler, using FFT, to unmask the weaker targets using three-dimensional PSR. 
\begin{table}[!b]
\centering
\caption{\footnotesize Comparison of execution times for 2D and 3D CLEAN on Zynq MPSoC}
\label{tab:CLEAN2Dvs3D}
{
\renewcommand{\arraystretch}{1.2}
\resizebox{\linewidth}{!}{
\begin{tabular}{|c|c|c|c|cc|}
\hline
\multirow{2}{*}{\textbf{CLEAN}} &
  \multirow{2}{*}{\textbf{\# packets}} &
  \multirow{2}{*}{\textbf{Doppler}} &
  \multirow{2}{*}{\textbf{Target}} &
  \multicolumn{2}{c|}{\textbf{Acceleration Factor}} \\ \cline{5-6} 
 &  &  &                    & \multicolumn{1}{c|}{\textbf{(MF, PSF, MUSIC/FFT)}}           & \textbf{Total}         \\ \hline
\multirow{3}{*}{\textbf{3D}} &
  \multirow{3}{*}{\textbf{100}} &
  \multirow{3}{*} {\textbf{\begin{tabular}[c]{@{}c@{}} 4096-FFT \\ (100,3996)\end{tabular}}}  &
  \multirow{1}{*}{3} &
  \multicolumn{1}{c|}{\multirow{1}{*}{1,1,1}} &
  \multirow{1}{*}{1}  \\ \cline{4-6}                                   &     &  &
  \multirow{1}{*}{2} & \multicolumn{1}{c|}{\multirow{1}{*}{1.5, 2, 1.5}}         & \multirow{1}{*}{1.57}                         \\ \cline{4-6} 
 &  &  & \multirow{1}{*}{1} & \multicolumn{1}{c|}{\multirow{1}{*}{3, NA ,3}}            & \multirow{1}{*}{3.65} 
 \\ \hline
\multirow{6}{*}{\textbf{2D}} &
  \multirow{6}{*}{\textbf{100}} &
  \multirow{3}{*}{\textbf{MUSIC}} &
  \multirow{1}{*}{3} &
  \multicolumn{1}{c|}{\multirow{1}{*}{2.94, 200.48, 0.09}} &
  \multirow{1}{*}{3.38}         \\ \cline{4-6} 
 &  &  & \multirow{1}{*}{2} & \multicolumn{1}{c|}{\multirow{1}{*}{2.97, 100.24, 0.136}} & \multirow{1}{*}{3.48} \\ \cline{4-6} 
 &  &  & \multirow{1}{*}{1} & \multicolumn{1}{c|}{\multirow{1}{*}{3, NA, 0.273}}        & \multirow{1}{*}{3.59} \\ \cline{3-6} 
 &
   &
  \multirow{3}{*} {\textbf{\begin{tabular}[c]{@{}c@{}} 4096-FFT \\ (100,3996)\end{tabular}}} 
&
  \multirow{1}{*}{3} &
  \multicolumn{1}{c|}{\multirow{1}{*}{2.94, 200.48, 1}} &
  \multirow{1}{*}{3.55}  \\ \cline{4-6} 
 &  &  & \multirow{1}{*}{2} & \multicolumn{1}{c|}{\multirow{1}{*}{2.97, 100.24, 1.5}} & \multirow{1}{*}{3.6} \\
  \cline{4-6} 
 &  &  & \multirow{1}{*}{1} & \multicolumn{1}{c|}{\multirow{1}{*}{3, NA, 3}}     & \multirow{1}{*}{3.65}  \\
  \hline
\end{tabular}}}
\end{table}
The execution times in terms of acceleration factor (AF) comparing both methods are reported in Table~\ref{tab:CLEAN2Dvs3D}. 
In the first row, 3D CLEAN utilizes only 100 packets for Doppler processing via FFT-based processing and subsequently, the data is zero-padded to 4096 slow time samples. Since detecting the third target using the 3D CLEAN with 100 packets is the most time-consuming, the corresponding AF is 1. For the second and first targets, further improvement in computational times is observed. For instance, matched filtering (MF) for the second and first targets is 1.5 and 3 times faster than MF for the third target, respectively. 3D CLEAN takes a longer time to generate the three-dimensional PSR versus the two-dimensional PSR in the 2D CLEAN. Hence, 2D CLEAN takes less computational time than 3D CLEAN and thus shows an AF of 3.4 for the third target over 3D CLEAN. Again, the gain in AF increases with the decrease in the number of targets. 
As a result, any possible performance improvement that may be realized by executing CLEAN on the radar data cube is offset by the cost incurred by the heavy computational time. This would ultimately result in longer radar signal processing and overall poorer communication metrics due to longer search times for optimal beam selection in ISAC system.
\color{black}
\section*{Appendix-II: Effect of Clutter}
We model static clutter as a collection of discrete scatterers spread throughout the radar field-of-view with strength varying based on an RCS drawn from an exponential distribution of $\sigma_{c_{avg}}$. Assuming we retain the same target statistics as discussed earlier, we consider a $\sigma_{c_{avg}} = 0.2 m^2$. As shown in Fig. \ref{fig:static_clutter_hist}, the distribution of the strength of the clutter scatterers has a mean clutter power of -87.3 dB, resulting in an SCR of -5.3 dB. Note that in these conditions, the noise is very low compared to clutter, and hence, it is more meaningful to discuss SCR rather than SNR. We identified approximately 820 clutter scatterers (along with just 3 targets) across multiple Monte Carlo simulation trials that are above the threshold. The resulting detection metrics implemented with 2D CLEAN and 3D CLEAN are provided in Table \ref{tab:clutter_missed_detections_3Dclean} for 200 Monte Carlo simulations.

\begin{table}[htbp]
\centering
\caption{\footnotesize Detection metrics of targets in the presence of discrete clutter scatterers with 3D and 2D CLEAN.}
\renewcommand{\arraystretch}{1.1}
\resizebox{0.75\linewidth}{!}{
\begin{tabular}{|l|l|l|}
\hline
\multirow{2}{*}{\textbf{Target Detection Metrics}}  &   \multicolumn{2}{c|}{\multirow{1}{*}{\textbf{(/200)}}}  \\ \cline{2-3}
& \textbf{3D CLEAN} & \textbf{2D CLEAN}\\ \hline
\multicolumn{1}{|c|}{T1: Missed Detections}  & \multicolumn{1}{c|}{0}   & \multicolumn{1}{c|}{0}   \\ \hline
\multicolumn{1}{|c|}{T2: Missed Detections} & \multicolumn{1}{c|}{45}  & \multicolumn{1}{c|}{46}    \\ \hline
\multicolumn{1}{|c|}{T3: Missed Detections} & \multicolumn{1}{c|}{107} & \multicolumn{1}{c|}{108}     \\ \hline
\multicolumn{1}{|c|}{False Alarms} & \multicolumn{1}{c|}{0} & \multicolumn{1}{c|}{62}     \\ \hline
\end{tabular}}
\label{tab:clutter_missed_detections_3Dclean}
\end{table}
The strongest target, T1, is almost always detected despite the presence of so many discrete clutter scatterers while the weaker targets - T2 and T3 - have more missed detections. The presence of a large number of static clutter scatterers results in fairly high false alarms with 2D CLEAN. Most of these false alarms arising from discrete static clutter scatterers are subsequently identified as distinct from the targets during the Doppler processing stage. Naturally, this issue does not arise with 3D CLEAN where the clutter scatterers are identified by their low Doppler values. The localization metrics for the three targets are shown in Table.\ref{tab:static_clutter_RMSE}. The error is greatest for the weakest target (T3), which is most affected by the presence of clutter compared to T2 and T1. 

\begin{table}[htbp]
\centering
\caption{\footnotesize RMSE of targets for SCR of -5.3 dB implemented with 2D CLEAN}
\renewcommand{\arraystretch}{1.1}{
\resizebox{0.65\columnwidth}{!}{
\begin{tabular}{|c|ccc|}
\hline
\multirow{3}{*}{\textbf{Target}} & \multicolumn{3}{c|}{\textbf{RMSE}}                                                                                                                                                                                                \\ \cline{2-4}
                 & \multicolumn{1}{c|}{\textbf{\begin{tabular}[c]{@{}c@{}}Range\\  (m)\end{tabular}}} & \multicolumn{1}{c|}{\textbf{\begin{tabular}[c]{@{}c@{}}Azimuth\\  (degrees)\end{tabular}}} & \textbf{\begin{tabular}[c]{@{}c@{}}Doppler velocity \\ (m/s)\end{tabular}} \\ \hline
1         & \multicolumn{1}{c|}{1.17}                                                 & \multicolumn{1}{c|}{2.20}                                                         & 2.10                                                              \\ \hline
2         & \multicolumn{1}{c|}{5.82}                                                 & \multicolumn{1}{c|}{11.43}                                                         & 6.21                                                              \\ \hline
3         & \multicolumn{1}{c|}{8.80}                                                 & \multicolumn{1}{c|}{13.80}                                                        & 10.91                                                             \\ \hline
\end{tabular}}}
\label{tab:static_clutter_RMSE}
\end{table}

Next, we consider the special cases where some of the static clutter scatterers occupy the same range-azimuth cells as the targets, as discussed in Section VI. This is where the super-resolution property of MUSIC becomes useful in distinguishing clutter from mobile users, specifically to decide if communication beams ought to be sent along those directions. Fig. \ref{fig:super_reso} shows the superiority of MUSIC over 4096-FFT(100,3996) in resolving multiple targets at the cost of increased hardware complexity as indicated in Table \ref{tab:musicVSfft}. Thus, users may consider adopting FFT in low-clutter environments (such as highways with smooth roads), while the more complex MUSIC may be adopted for high-clutter environments (such as urban environments or roads with high surface roughness or the presence of tunnels and over-bridges).
\section*{Appendix-III: Coherent Integration}
\begin{figure*}[!ht]
    \centering
    \includegraphics[scale = 0.4]{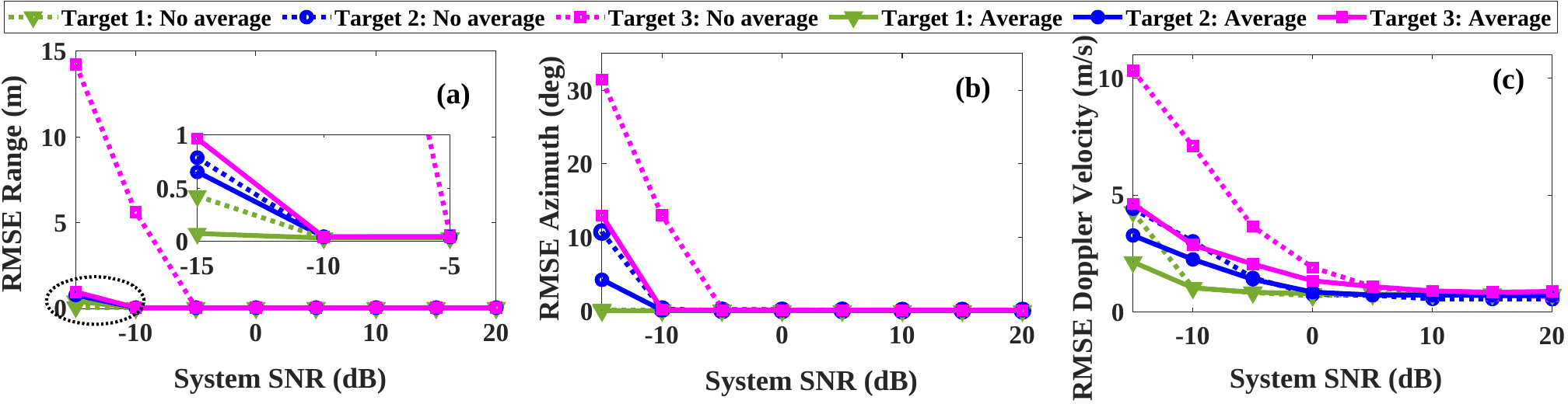}
    \vspace{-0.2cm}
    \caption{\small Averaged range, azimuth, and Doppler velocity RMSE across 10 packets for three targets with LOS on MPSoC. Based on strength, targets 1,2 and 3 have an offset of 39 dB, 22.5 dB, and 12.6 dB compared to system SNR.}
    \label{fig:avg_rmse_los}
\end{figure*}

\begin{figure*}[!ht]
    \centering
    \includegraphics[scale = 0.4]{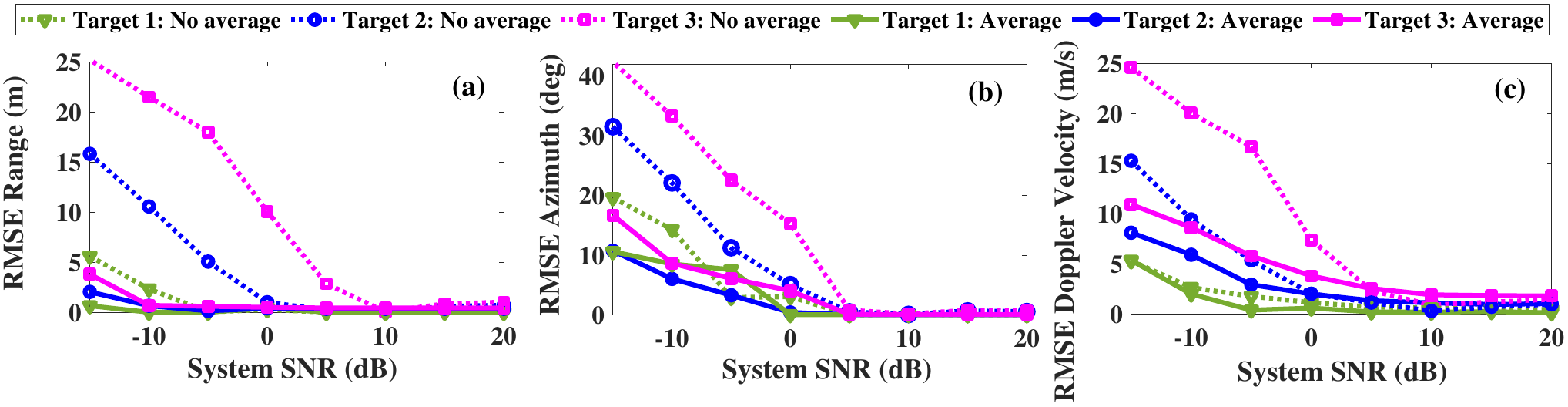}
    \vspace{-0.3cm}
    \caption{\small Averaged range, azimuth, and Doppler velocity RMSE across 10 packets for three targets with Rician channel on MPSoC. Based on strength, targets 1,2 and 3 have an offset of 39 dB, 22.5 dB, and 12.6 dB compared to system SNR.}
    \label{fig:avg_rmse_rician}
\end{figure*}

We perform averaging operations across multiple packets. Here, the returns from the moving target add up coherently, while the clutter returns integrate non-coherently, resulting in a superior signal-to-clutter ratio after integration. The RMSE of the range, azimuth, and Doppler estimates after integration are shown in Fig.~\ref{fig:avg_rmse_los} and Fig.~\ref{fig:avg_rmse_rician}  for LOS and Rician channels, respectively. A significant reduction in RMSE is observed after the averaging operation. Since the averaging operation is performed on PS, additional FPGA resources are not needed. However, the total execution time increases with the increase in the number of packets used for averaging operations. 

\bibliographystyle{ieeetr}
\bibliography{references}

\end{document}